\documentstyle[preprint,aps,epsfig]{revtex}
\def\beq{\begin{equation}}
\def\eeq{\end{equation}}
\def\beqa{\begin{eqnarray}}
\def\eeqa{\end{eqnarray}}
\def\lb{\label}

\def\GeV{\nobreak\,\mbox{GeV}}

\def\Tr{\mbox{ Tr }}
\def\La{\Lambda}
\def\ka{\kappa}
\def\la{\lambda}
\def\ga{\gamma}
\def\om{\omega}
\def\nn{\nonumber}
\begin{document}
\draft

\title{ Form Factors and Decay Rates for  Heavy $\Lambda$ Semileptonic 
Decays from QCD Sum Rules }

\author { R.S. Marques de Carvalho$^1$,  
F.S. Navarra$^1$, M. Nielsen$^1$, E. Ferreira$^2$ and H.G. Dosch$^3$}
\address{  $^1$Instituto de F\'{\i}sica, Universidade de S\~{a}o Paulo\\
  C.P. 66318,  05315-970 S\~{a}o Paulo, SP, Brazil
\\[0.1cm]
 $^2$Instituto de F\'{\i}sica, Universidade Federal do Rio de  
Janeiro \\
C.P. 68528, Rio de Janeiro 21945-970, RJ, Brazil\\
 [0.1cm]
 $^3$Institut f\"ur Theoretische Physik, Universit\"at Heidelberg\\
 Philosophenweg 16, D-6900 Heidelberg, Germany}

\maketitle

 \begin{abstract}

QCD sum rules for the determination of form factors of $\La_b$ and 
$\La_c$ semileptonic decays are investigated. With  a form for the
baryonic current appropriate for the limits of the heavy quark 
symmetries, the different tensor structures occurring in the 
two- and three-point functions are separately studied, and in each case 
general relations are written  for the form factors. 
Particular 
attention is given to the treatment of the kinematical region ascribed 
to the continuum. The $t$-dependence of the form factors and the decay 
rates are numerically evaluated and compared to experimental information. 

 \end{abstract}  

\bigskip
PACS Numbers~ :~ 12.38.Lg , 13.30.Ce , 14.65.Dw , 14.65.Fy

\bigskip

\bigskip

{\bf 1. Introduction }

In this paper we treat semileptonic decays of the heavy baryons 
$\Lambda_c$ and $\Lambda_b$.
The study of the matrix elements of weak decays of heavy hadrons
are an important source of information on some 
elements of the Cabbibo-Kobayashi-Maskawa (CKM) matrix, but
the determination of these fundamental quantities of the standard model 
require disentanglement from the effects of strong interactions 
occurring inside the hadrons.

QCD sum rules \cite{SVZ79} (for reviews see 
\cite{Nar89,Shif92}) provide an important approach for calculating
hadronic matrix elements and form factors for systems with both light
and heavy quarks. This method  deals with  the nonperturbative 
aspects of QCD analytically using a limited input of phenomenological 
parameters and has been applied  successfully to semileptonic decays 
of mesons.  
Unfortunately QCD sum rules are not expected  to do as well for baryonic 
as for mesonic amplitudes. This difficulty is inherent to the
method, as the basis of the sum-rule approach is an expansion in 
{\em local} operators. As a starting point the hadron is represented by a 
local interpolating field constructed from quarks. It is
evident that for the case of baryons where at least three operators are
taken at the same point this is a more drastic reduction than for the
mesonic case with two operators. We therefore expect the contribution
of higher resonances  to be even more important for baryons than for
mesons. Technically this reflects itself in the higher dimension of the
interpolating fields and thus in a faster increase of the perturbative
contribution, which makes the assumptions on the continuum contributions
more decisive.

In the limit where one of  the quarks in the initial and final hadrons 
is infinitely heavy there are new flavor and spin symmetries,
which have received much attention in the last years in the framework 
of  the so-called Heavy Quark Effective Theory (HQET) 
\cite{Shu82,VolSh87,NusW87,iw90,g90,ma91} (for a review see \cite{n94}).
QCD sum rules simplify considerably in the heavy quark limit
 \cite{Shu82}, but they can be applied also to evaluate 1/M corrections 
to the results obtained with infinite heavy quark masses.
The method has had good  success in the calculation of corrections to HQET
(see \cite{n94} and references therein).

In the present paper, which is an extended and revised  version of a 
previous letter \cite{dfnr98}   we evaluate the semileptonic decays of heavy 
$\La$-baryons in the QCD sum rule approach as it was developed a few years 
ago for heavy meson decays \cite{BBD91}.  This approach treats full QCD
but also reproduces the symmetries of HQET. In $\La_c$ 
semileptonic decay, which is among the best investigated processes of 
this kind, the  relevant non-Cabbibo suppressed CKM matrix element $V_{cs}$ is 
known, so that calculations of these decays may provide very good 
tests  of the applied method. On the other hand, there are quite serious 
discrepancies between experiments and HQET in the ratios of the lifetimes 
of beautiful baryons to beautiful mesons 
(see e.g. \cite{neu97}), and hence it is of prime interest to investigate 
in full QCD all decay channels of the $\La_b$ baryon.

We use the sum rule technique for three-point functions that was introduced
\cite{ios82,ne82} for the study of the pion eletromagnetic form factor 
at intermediate Euclidean momentum transfer. We analyse our form factors
in the physical region for positive values of the momentum transfer since
in this case the cut in the $t$ channel starts at $t\sim m_Q^2$ ($m_Q$ 
represents the heavy quark mass), and thus the Euclidean region stretches 
up to that threshold.

This paper is organized as follows: in Sec. II we discuss the method and 
introduce the form factors to be evaluated. In Sec.~III we collect and 
discuss the results for the two- and three-point functions sum rules in HQET 
and shortly discuss the results. In Sec. IV we present our results obtained 
from QCD sum rules for the form factors and decay rates for $\Lambda_c$ 
and $\Lambda_b$ decays, and finally in Sec. V we present our conclusions. 
Technical details about the method used are presented in Appendices A, B 
and C.

\bigskip

{\bf 2. Sum Rule Calculation }

We label generically the initial channel by $I$ and the final 
channel by $F$. The weak transition current couples initial quark $Q_I$ 
(mass $m_I$) to final quark $Q_F$ (mass $m_F$), and
we assume that the other two non-interacting quarks are massless 
($ m_u= m_d=0$.). The initial and final baryon four-momenta are called 
$p_I$ and $p_F$, with corresponding Mandelstam variables $s=p_I^2$ and 
$u=p_F^2$ respectively. The current four-momentum is called 
$q= p_I-p_F$, and 
$t=q^2$. We also define the four-vector $r=p_I+p_F$.
The initial and final baryons
are called $\Lambda_I$ and $\Lambda_F$, and their masses are 
$M_{\La_I}$ and $M_{\La_F}$. 

In order to study the decay $\Lambda_I \to \Lambda_F + \ell + \nu_\ell$
using the QCD sum rule approach we consider the three-point function
of the weak transition  current from a initial to a final quark  
      $J_\mu= \bar Q_F \gamma_\mu (1- \gamma_5) Q_I$ 
and the interpolating fields of the initial and final baryons 
$\eta_{\Lambda_I}$ and $\eta_{\Lambda_F}$
\begin{equation}
\Pi _\mu (p_F,p_I)=i^2\int d^4xd^4y ~ \langle 0|T\{\eta _{\Lambda_F}(x)
J_\mu (0)\overline{\eta }_{\Lambda_I}(y)\}|0\rangle ~  
e^{ip_Fx}e^{-ip_Iy}\; . 
\label{cor}
\end{equation}

This three-point function of two spinors and a vector and axial vector can 
be expressed by a superposition of scalar form factors and the 24 vector
valued 4$\times$4 matrices of
Table 1.

The method of sum rules is based on the simultaneous evaluation 
of the expression of the correlator (\ref{cor}) in a phenomenological
representation obtained by inserting physical intermediate states
(the usual spin $\frac{1}{2}$ baryons and  higher 
resonances), and in a theoretical representation obtained through
the OPE expansion. This is detailed below. 

\subsection{Phenomenological Side}

The phenomenological expression is continued by dispersion relations into 
the not-so-deep 
Euclidean region where $p_F^2<M_{\La_F}^2$ and $p_I^2<M_{\La_I}^2$ and where 
it is believed that the theoretical representation is reliable.
We introduce the couplings $f_{I}$ and $f_{F}$ of the currents with 
the respective spin $\frac{1}{2} $  hadronic states 
\begin{eqnarray}
\langle 0|\eta _{\Lambda_F}|\Lambda_F(p_F)\rangle &=&f_{F} u_F(p_F)~ ,
\label{lam}
\\
\langle\Lambda_I(p_I)|\overline\eta _{\Lambda_I}|0\rangle &=& f_{I}
\overline{u}_I(p_I)\; ,
\label{lamc}
\end{eqnarray}
and obtain the phenomenological representation of Eq. (\ref{cor})
\begin{eqnarray}
\label{phen1}
\Pi_\mu^{(\rm phen)}(p_F,p_I) &=&\sum_{\rm spins}
{ {\langle 0|\eta _{\Lambda_F}|\Lambda_F(p_F)\rangle}
 \over{p_F^2-M_{\La_F}^2}}
\langle\Lambda_F(p_F)|J_\mu|\Lambda_I(p_I)\rangle
{ {\langle\Lambda_I(p_I)|\overline\eta _{\Lambda_I}|0\rangle}
\over{p_I^2-M_{\La_I}^2}}   \nonumber \\
 &+&  \mbox{higher resonances}  \nonumber\\ 
&=&\sum_{\rm spins}
{     { f_F ~ u_F(p_F) }
 \over{p_F^2-M_{\La_F}^2}}
\langle\Lambda_F(p_F)|J_\mu|\Lambda_I(p_I)\rangle
{ {  f_I ~ \overline u_I(p_I)  }
\over{p_I^2-M_{\La_I}^2}}   \nonumber \\
 &+&  \mbox{higher resonances}  \; .
\end{eqnarray}

The general transition element $\langle\Lambda_F(p_F)|J_\mu|\Lambda_I(p_I)\rangle$
with a weak  current between two spin $\frac{1}{2}$ states
can be written  in terms of the moments $q=p_I-p_F$ and $r=p_I+p_F$,
the spinors $\overline u_F(p_F)$ and $ u_I(p_I)$ and the 
24 structures made with $\gamma $ matrices (12 vector $v_\mu^i$ 
and 12 axial vector $a_\mu^i $ quantities) listed in Table I in 
the form  
\begin{equation}
\langle\Lambda_F(p_F)|J_\mu|\Lambda_I(p_I)\rangle= 
     \overline u_F(p_F)
\sum_1^{12} ~ \bigg[ f_i^V~  v_\mu^i~ + ~f_i^A~ a_\mu^i  \bigg]
               u_I(p_I)~.
\label{curr1}
\end{equation}

\vskip1.0cm
\begin{center}
{\bf Table I}

{\small Vector and axial vector structures constructed with 
four-vectors $q$, $r$ and \\
combinations of $ \gamma$ matrices.
Notation is from Bjorken-Drell, with $\sigma_{\mu \nu}
=\frac{i}{2}(\gamma_\mu \gamma_\nu-\gamma_\nu \gamma_\mu) $.}

\vskip0.5cm

\begin{tabular}{ c c c }  \hline
i & $v^i_\mu$ & $a^i_\mu$ \\
 \hline \hline 
 1 & $q_{\mu }\,1$ & $q_{\mu }\,\gamma_{5}$  \cr 
 2 & $r_{\mu }\,1$ &  $ r_{\mu }\,\gamma_{5}$  \cr 
 3 &  $\gamma_{\mu }$  &  $\gamma_{\mu } \gamma_{5}$    \cr
 4 & $q_{\mu }\, \not\! q$  & $q_{\mu }\, \not\! q \gamma_{5}$  \cr 
 5 &   $r_{\mu }\, \not\! q$  & $r_{\mu }\, \not\! q \gamma_{5}$  \cr 
 6 &   $q_{\mu }\, \not\! r$  & $q_{\mu }\, \not\! r \gamma_{5}$  \cr 
 7 &   $r_{\mu }\, \not\! r$  & $r_{\mu }\, \not\! r \gamma_{5}$  \cr 
 8 & $i\sigma_{\alpha\mu}q^\alpha$ & $i\sigma_{\alpha\mu}q^\alpha 
\gamma_5$ \cr 
 9 & $i\sigma_{\alpha\mu}r^\alpha$ & $i\sigma_{\alpha\mu}r^\alpha \gamma_5$ \cr
10 &  $  i \sigma_{\alpha\beta}q^\alpha r^\beta q_\mu$ &
      $i \sigma_{\alpha\beta}q^\alpha r^\beta \gamma_5 q_\mu$  \cr
11 &  $i \sigma_{\alpha\beta}q^\alpha r^\beta r_\mu$ &
      $ ~ i \sigma_{\alpha\beta}q^\alpha r^\beta \gamma_5 r_\mu$ ~  \cr
 12 & $~ \epsilon_{\rho\beta\alpha\mu} q^\rho r^\beta\gamma^\alpha\gamma_5 ~ $
     & $\epsilon_{\rho \beta \alpha \mu} q^\rho r^\beta \gamma^\alpha$ \cr 
\hline
\end{tabular}
\end{center}

\bigskip

  Since on-mass-shell $\Lambda_I$ and $\Lambda_F $ baryons obey 
Dirac equations
 
we may use the projector properties of the sums over the (on-mass-shell)
spinors
\begin{equation}
\sum_{\rm spins} u_I(p_I)\overline u_I(p_I)=\gamma \cdot p_I + M_{\La_I}~,
~ \sum_{\rm spins} u_F(p_F)\overline u_F(p_F)=\gamma \cdot p_F + M_{\La_F}~,
\label{dirac2}
\end{equation}
and the sum over 24 invariants in Eq. (\ref{curr1}) can be reduced to only 
6 independent terms. 

The experimental information is usually represented \cite{Kor94}
by the invariant decomposition of the amplitude in the form 
\begin{eqnarray}
\langle\Lambda_F(p_F)|J_\mu|\Lambda_I(p_I)\rangle&=& \nonumber\\
&&\hspace{-3cm}\overline{u}(p_F)
[\gamma_\mu
(F_1^V+F_1^A\gamma_5)+i\sigma_{\mu\nu}q^\nu(F_2^V+F_2^A\gamma_5)+
q_\mu (F_3^V+F_3^A\gamma_5)]u(p_I)\; ,  
 \label{ff} 
\end{eqnarray}
where the form factors $F^{V,A}_i ~ ~(i=1,2,3)$  are functions of 
$t=q^2$ .

We then obtain for the phenomenological representation of the 
correlator
\begin{eqnarray}
\Pi_\mu^{(\rm phen)}(p_F,p_I) &=&{(f_{F}\rlap{/}{p_F}+f'_{F})
\over p_F^2-
M_{\La_F}^2} ~  \bigg[\gamma_\mu
(F_1^V+F_1^A \gamma_5)+i\sigma_{\mu\nu}q^\nu(F_2^V+F_2^A\gamma_5)
\nonumber \\*[7.2pt]
&+ & 
q_\mu (F_3^V+F_3^A \gamma_5)\bigg]~ {(f_{I}\rlap{/}{p_I}+f'_{I})
 \over p_I^2-M_{\La_I}^2} + \mbox{higher resonances}\; ,
\label{phen2}
\end{eqnarray}
where we have defined 
\begin{equation} 
\label{coupling}
  f'_F=f_{F}M_{\La_F} ~  ~ , ~ ~  f'_{I}=f_{I}M_{\La_I}~ .
\end{equation}
The reason for the separate notations for $f$ and $f'$ is that 
they belong to terms with different spin content (with and without 
$\gamma$  matrix) and this will help in the classification of 
types of traces to be evaluated. Another reason is that we shall
determine $f$ and $f'$ from  the mass sum rules for $\La_I$ and
$\La_F$, and there is no guarantee that we have exactly  
$f'_I=f_I M_{\La_I}$,  $f'_F= f_F M_{\La_F} $ .

Referring to the spinor forms that appear in Eq. (\ref{phen2}),
we may pick up terms with or without $\gamma$ matrices in the 
numerators of the two fractions, and thus form four different kinds
of products that we identify as 
\begin{equation}
\label{unk}
F^{V,A}_i(q^2) f_I f_F ~ ,~  F^{V,A}_i(q^2) f_I f'_F~  ,~ 
F^{V,A}_i(q^2) f'_I f_F ~  ,~  F^{V,A}_i(q^2) f'_I f'_F  ~ , ~  
~ ({\rm with}~  i=1,2,3) ~. 
\end{equation}
We project out a sum rule for each one of these 24 products
by performing  appropriate traces using  Eq. (\ref{phen2}). 
Independent traces can be formed after multiplying Eq. (\ref{phen2}) by 
the 12 vector and 12 axial-vector structures of  Table I
 \beq \lb{3.1}
x_i \equiv \Tr[v^i_\mu \Pi^{\mu,(\rm phen)}] \mbox{~  and  ~ } 
 u_i \equiv \Tr[a^i_\mu \Pi^{\mu,(\rm phen)}], \quad i=1,\dots 12 ~,
\label{xu}
\eeq
 so that we 
obtain  $2\times 12$ independent relations  that must be solved 
for the $2 \times 12$ unknowns quantities mentioned in Eq. (\ref{unk}) 
(we must recall that before  we must introduce $q$ and $r$ everywhere 
in the  places of $p_I$, $p_F$). The results are given in Appendix A.

The constants $f_I$, $f'_I$, $f_F$, and $f'_F$ have to be calculated from 
two-point function sum rules, as usual.

\subsection{Theoretical Side}

The theoretical counterpart is evaluated by performing the Wilson's 
operator product expansion (OPE) of the operator in Eq.~(\ref{cor}) and 
then taking the expectation value with respect to the physical vacuum. The 
term from 
the unit operator gives the usual perturbative contribution, while  the 
vacuum expectation values of the other operators in the expansion give 
the nonperturbative corrections proportional to the condensates of the 
respective operators. Thus 
\beq
\Pi^{(\rm theor)}_\mu = \Pi^{(\rm pert)}_\mu + 
      \sum_i \Pi^{({\rm nonpert}(i))}_\mu ~ ,
\label{theor1}
\eeq
where the index $i$ refers to the types and dimensions of the 
condensates.

As usual in the sum rule method,  we wish to  evaluate the form 
factors that build the $\langle \Lambda_F |J_\mu| \Lambda_I\rangle$ 
 amplitude by matching the phenomenological 
representation Eq. (\ref{phen2}) of the three point function with 
its theoretical counterpart from  Eq. (\ref{theor1}). The higher 
resonances are approximated by the perturbative contribution 
invoking quark hadron duality \cite{SVZ79}. 

 As it is well known from two-point sum rules for baryons
\cite{Iof81,CDKS82,bcdn92},  there is a continuum
 of choices for the interpolating currents. Of course the results should in 
principle be independent of the choice of the current (except for 
pathological cases which couple very weakly to the ground state), but the 
justification of the approximations depends on the choice made. From the point
of view of heavy quark symmetries, the spin of the heavy $\La$ is carried
by the heavy quark, with the light quarks being in a spin and isospin singlet 
state, namely 
\beq
\eta_{\La_Q} = \epsilon_{ABC}(\bar u^A \gamma^5 d^B) Q^C ~ ,
\label{curr}
\eeq
where $u^A$ and $d^B$ stand for the Dirac fields of light quarks of colors 
$A$ and $B$, and $Q^C$ represents  a heavy quark ($b$ or $c$) of color $C$. 
This current couples strongly to the $\Lambda$ states in the heavy quark 
limit, and with this choice the quark condensate and the mixed quark-gluon
condensate do not contribute to Eq. (\ref{theor1}). The gluon condensate does 
contribute, but experience with baryonic two-point functions and mesonic 
three point functions teaches us that it is of little influence. We are thus 
left with only the perturbative and the four quark condensate contributions.

 Once we have calculated explicitly  the terms contributing  to
$\Pi_\mu^{(\rm theor) }$ we can form the traces

 \beq 
x_i \equiv \Tr[v^i_\mu \Pi^{\mu,{(\rm theor)}}] \mbox{~  and ~  } 
 u_i \equiv \Tr[a^i_\mu \Pi^{\mu,{(\rm theor)}}], \quad i=1,\dots 12 ~,
\label{tr}
\eeq
and thus project out the form factors given in Appendix A.

\subsection{Sum Rule}

To the 24 invariants in Eq. (\ref{curr1}) correspond in the off-mass-shell
case 24 invariant functions $F_i^{V,A}(s,u,t)f_I^{(')} f_F^{(')}$ 
(where by $f_I^{(')}$  we mean either $f_I$ or $f'_I$, and similarly 
for  $ f_F^{(')}$ ) given in Appendix A. They obey double dispersion  
relations of the form
\begin{equation}
F_i^{V,A}(s,u,t)f_I^{(')} f_F^{(')} = {1\over\pi^2}\int_{m_I^2}^\infty ds'
\int_{m_F^2}^\infty du'
{\rho_{i}^{V,A}(s',u',t)\over (s'-s)(u'-u)} + \cdot\cdot\cdot \; ,
\label{spec}
\end{equation}
where $-4\rho_{i}^{V,A}(s,u,t)$ equals the double discontinuity of the
amplitude $F_i^{V,A}(s,u,t)f_I^{(')} f_F^{(')}$ on the cuts $m_I^2\leq s<
\infty$, $m_F^2\leq u<\infty$, which is given by the double discontinuity
of the right hand side of the expressions in Appendix A. In Eq.~(\ref{spec})
$m_{I}$ and $m_{F}$ are respectively the  initial and final heavy quark 
masses  and
the dots represent subtraction polynomials in $s$ and 
$u$, which will vanish under the double Borel transform \cite{ios82},
in a straightforward generalization of that used before \cite{SVZ79}.
Applying the double Borel transform to Eq.(\ref{spec}) and subtracting the
continuum contribution we obtain 
\beqa
F_i^{V,A}(t)f_I^{(')} f_F^{(')}e^{-M_{\Lambda_I}^2/M_I^2} e^{-M_{\Lambda_F}
^2/M_F^2} 
&=& {1\over\pi^2}\int_{m_I^2}^{\infty} ds'
\int_{m_F^2}^{\infty} du' R_{cont}(s',u',s_0,u_0)
\nonumber \\*[7.2pt]
&& \times\rho_{i}^{V,A}(s',u',t)e^{-s'/M_I^2} e^{-u'/M_F^2} \; ,
\label{specb}
\eeqa
where $M_I^2$ and $M_F^2$ are the Borel parameters and 
$R_{cont}(s',u',s_0,u_0)$ defines the continuum model with
$s_0$ and $u_0$ being,
respectively, the continuum thresholds for the $\Lambda_I$ and 
$\Lambda_F$ baryons, determined from the mass sum rules. 

We will consider 
two different region of integration for the continuum contribution. A
rectangular region specified by a rectangular cutoff:
\beq
R_{cont}(s',u',s_0,u_0)= \Theta(s_0-s')\Theta(u_0-u') \; ,
\label{contquad}
\eeq
and a triangular region determined by:
\beq
R_{cont}(s',u',s_0,u_0)= \Theta\left(u_0 - \frac{m_F}{m_I}(s'-s_0) -u'\right)
\; .\label{conttri}
\eeq

In the OPE side of the sum rules we can evaluate the double discontinuity
of the traces $x_i$  using  Cutkosky's rules. The perturbative 
contribution to the double discontinuity of $x_i$ in  Eq.(\ref{tr}) 
is given by
\beqa
{\cal DD}[x_i]_{\rm (pert)}&=&-{6\over(2\pi)^3}{1\over8\sqrt{\lambda(s,u,t)}}
\int_0^{(\sqrt{s}-m_I)^2}
dm^2\; m^2  \int dK_0d{|\vec K|}^2d\cos\theta_Kd\phi_K 
\nonumber \\*[7.2pt]
&& \times \delta(K_0-\overline{K}_0)
\delta(|\vec{K}|^2-\overline{|\vec{K}|^2})\delta(\cos\theta_K-
\overline{\cos\theta_K})\Theta(1-\overline{\cos\theta_K}^2)N_i\; ,
\label{per}
\eeqa
where  $$\overline{K}_0=(s+m^2-m_I^2)/2\sqrt{s} ~ ~  ,~ ~ 
       \overline{|\vec{K}|^2}= \lambda(s,m^2,m_I^2)/4s $$ 
and
\beq
\overline{\cos\theta_K}={2p_{s_0}\overline{K}_0-u-m^2+m_F^2\over
2|\vec{p_s}||\vec{K}|}\; ,
\eeq
where 
$$ p_{s_0}=(s+u-t)/2\sqrt{s} ~ ,~  |\vec{p_s}|^2=\lambda(s,u,t)/4s $$ 
and
$$\lambda(x,y,z)=x^2+y^2+z^2-2xy-2xz-2yz ~.$$
In Eq.(\ref{per}) $N_i$ represents the traces of the Lorentz structure
of the three-point function multiplied by each of the 12 independent 
vector structures $v_i^\mu$ 
\beq
N_i=\Tr[(\rlap{/}{p_s}-\rlap{/}{K}+m_F)\gamma_\mu(1-\gamma_5)(\rlap{/}{p_c}
-\rlap{/}{K}+m_I) v_i^\mu]\; .
\eeq
There are, of course, equivalent expressions for the double discontinuity of
$u_i$ (see Eq.(\ref{tr})) where the Lorentz structure is multiplied 
by the axial vector quantities  $a_i^\mu$.

In order to estimate the four-quark condensate we use the factorization
\beq
\langle \bar{d}^A_\alpha\bar{u}^B_\beta u^{B'}_{\beta'}d^{A'}_{\alpha'} 
\rangle = {\kappa\over12^2} ~ \delta_{\beta\beta'}\delta_{\alpha\alpha'}
\delta^{AA'}\delta^{BB''} \langle\bar{u}u\rangle\langle\bar{d}d\rangle ~ ,
\label{cond}
\eeq
where the parameter $\kappa$, which may have some value  from $1$ to $3$, 
is introduced 
to account for deviations from the factorization hypothesis \cite{Nar89}. 

For the particular choice of current shown in Eq.~(\ref{curr}) 
and using only the sum rules based on $f_I$ and $f_F$ structures (for which
the imaginary parts are positive definite) we obtain the general results
\beq 
\label{general}
F^A_1(t) =-\, F^V_1(t)~ ; \; F^V_2(t)= F^V_3(t)= F^A_2(t)= F^A_3(t) = 0\; ,
\eeq
with the $t$-dependence given by 
\beqa
F_1^V(t) &=& {\frac{e^{M_{\Lambda_I}^2/M_I^2} e^{M_{\Lambda_F}^2/M_F^2}}
      { 20\pi^4f_I f_F }} \bigg[ {15 \over 16}\int_{m_I^2}^{\infty} ds'
\int_{m_F^2}^{\infty} du' { \frac{P(s',u',t)} {[\lambda(s',u',t)]^{5/2}} }
   e^{-s'/M_I^2} e^{-u'/M_F^2} 
    \nonumber  \\
&&  \Theta(u'-u_{\rm min}) \Theta(u_{\rm max}-u')\;
R_{cont}(s',u',s_0,u_0)
+{10\over3}\pi^4\kappa \langle\overline{q}q\rangle^2e^{-m_I^2/M_I^2} 
e^{-m_F^2/M_F^2}  \bigg] ~ ,
\label{f1v}
\eeqa
where
\beqa
P(s,u,t)={1\over2} \{ m_I^4u+s[m_F^4+m_F^2(s-t-u)+tu]-m_I^2[(s+t-u)u 
% \nonumber \\*[7.2pt]
  + m_F^2(s-t+u)] \}^2 ~ ,
\label{p}
\eeqa
and the quantities
\beq
u_{\tiny{\begin{array}{l}
{\rm min}\\{\rm max}\end{array}}}={1\over2m_F^2}\left\{m_F^2(s+m_I^2)+
(s-m_I^2)\left[
m_I^2-t\mp\sqrt{\lambda(m_I^2,m_F^2,t)}\right]\right\}\; ,
\label{umima}
\eeq
define the physical region.

As mentioned before, $f_I$ and $f_F$ are determined from the two point sum 
rules and are given by \cite{bcdn92}
\beqa
f_{I(F)}^2&=&{e^{M_{\Lambda_{I(F)}}^2/M_{MI(MF)}^2}\over20\pi^4}\left[
{5\over128}m_{I(F)}^4\int_{m_{I(F)}^2}^{s_0(u_0)}ds~  
     h(s)~ e^{-s/M_{MI(MF)}^2}
\right.\nn\\
&+&{5\over768}\langle g_s^2G^2\rangle\int_{m_{I(F)}^2}^{s_0(u_0)}ds~ 
 (1-x)(1+5x^2)~ e^{-s/M_{MI(MF)}^2}+
\left.{10\over3}\pi^4\kappa\langle\overline{q}q\rangle^2
e^{-m_{I(F)}^2/M_{MI(MF)}^2}\right]\; ,
\label{fIF}
\eeqa
with
\beq
h(s)=(1-x^2)\left(1-{8\over x}+{1\over x^2}\right)-12\ln{x}\; ,
\label{hs}
\eeq
where $$x=m_{I(F)}^2/s~ ,$$  while $M_{MI(MF)}$ represent the Borel
masses in the two-point functions of $\La_{I}$ and $\La_{F}$ 
respectively.
In Eq.~(\ref{fIF}) we have included the gluon condensate, 
$\langle g_s^2G^2\rangle$, contribution for completeness, but  
its contribution is negligible and does not influence meaningfully 
 the actual calculation. 

Comparing Eqs.~(\ref{f1v}) and (\ref{fIF})
we can see that the exponentials multiplying Eq.~(\ref{f1v}) 
disappear 
if we choose 
\begin{equation}
         2M_{MI(MF)}^2=M_{I(F)}^2 ~ .
\label{borel}
\end{equation}
 Indeed, this way of relating 
the Borel parameters in the two- and three-point functions is a crucial
ingredient for the incorporation of the HQET symmetries, and leads
to a considerable reduction of the sensitivity to input parameters, 
such as continuum thresholds $s_0$ and $u_0$, and to radiative 
corrections \cite{BBG93}.

\bigskip

{\bf 3. Leading HQ limit }

In the HQET limit for infinite quark masses, the expressions for the
sum rules simplify considerably, as was first noted by 
Shuryak\cite{Shu82}.  All form factors can be described in terms of a
simple function, the Isgur-Wise (IW) function, $\xi$, normalized to 1 at 
maximum momentum transfer. We use the familiar $y$-variable, related to
the square of momentum transfer by
\beq \label{t}
t = m_I^2 + m_F^2 - 2 m_I\, m_F ~  y ~ .
\eeq
For convenience we give the relation between the Dirac-type form factors
used here and the velocity form factors appropriate for  HQET. The latter are
defined through
\beqa
\langle\La_F(p_F)|J_\mu|\La_I(p_I)\rangle&=& \nonumber\\
&&\hspace{-3cm}
\bar u(p_F) \left( \gamma_\mu G^V_1 + v_\mu^I G^V_2 + v_\mu^F G^V_3
+\ga_5 (\gamma_\mu G^A_1 + v_\mu^I G^A_2 + v_\mu^F G^A_3 )\right) u(p_I) ~ ,
\eeqa
where $v^I$ and $v^F$ are the velocities of the initial and final
hadron respectively, and the form factors are functions of $y$.

Neglecting the matching conditions one has
$G^V_1\, = \, G^A_1 \, = \xi(y) + O\left(\frac{1}{m}\right)$, and all 
other form factors are  $O\left(\frac{1}{m}\right)$ ~ . 
The relations with the form factors defined in Eq.~(\ref{ff}) are given by
\cite{n94,Kor94}

\beqa
F^V_1(t) &=& G^V_1 +(m_F+m_I)\left(\frac{1}{2 m_I} G^V_2+\frac{1}{2 m_F}
G^V_3\right) ~ ,  \nn\\
F^V_2(t) &=& -\frac{1}{2 m_I} G^V_2-\frac{1}{2 m_F} G^V_3 ~ , \nn \\
F^V_3(t) &=& -\frac{1}{2 m_I} G^V_2+\frac{1}{2 m_F} G^V_3  ~ ,\nn \\
F^A_1(t) &=& -G^A_1 -(m_F-m_I)\left(\frac{1}{2 m_I} G^A_2+\frac{1}{2 m_F}
G^A_3\nn \right) ~ ,  \\
F^A_2(t) &=& \frac{1}{2 m_I} G^A_2+\frac{1}{2 m_F} G^A_3 ~ ,  \nn \\
F^A_3(t) &=& \frac{1}{2 m_I} G^A_2-\frac{1}{2 m_F} G^A_3 ~ . 
\label{FG}
\eeqa

The baryonic two-point functions were
discussed  both in full QCD and in the HQ-limit in ref.
\cite{bcdn92}, three-point functions in ref. \cite{GroY92}, and
1/m corrections have been considered in ref. \cite{Dai96}.

In this section we collect and discuss the results for the two-  and 
three-point sum rules. We  stay on the firm grounds of
Wilson's OPE and only use local condensates. The four-quark condensate,
which is the most important non-perturbative input, is normally reduced 
to the square of the chiral symmetry
breaking two-quark condensate through the factorization hypothesis.
We allow for an uncertainty in the factorization  hypothesis introducing 
a factor $\ka$ that takes some value between 1 and 3 (see Eq.(\ref{cond})).

Usually one  introduces the variables $\om$ and $\mu$ related to the 
relativistic momentum square $s$  and the variable $M^2$ which appears
after Borel improvement by
\beq
s=m^2 + 2 m \om ~ ~  ; \qquad  M^2 = 2 m \mu ~  ,
\label{sm}
\eeq
where $m$ represents the heavy quark mass ($m_I$ or $m_F$).
Then the square of the coupling to the $\La$ current ($\rlap{/}{q}$-
structure) is given by the sum rule derived from the two-point function
\beq \label{twopoint}
f^2 = \exp\left[\frac{M_{\La}^2-m^2}{2 m \mu} \right]\frac{1}{20 \pi^4}
\left(I_5(\mu) 
+ \frac{5}{32} \langle g_s^2 G^2\rangle I_1(\mu)
+ \frac{10}{3} \pi^4 \ka\langle\bar q q\rangle^2  \right)
+O\left(\frac{1}{m}\right) \; ,
\eeq
with 
\beq
I_n(\mu)=\int_0^{\om_c} \om^n e^{-\om/\mu} d\om  ~  .
\eeq
Here the cutoff $\om_c$ is the continuum threshold, {\it i.e}, above this 
value the 
resonances are taken into account by perturbation theory. One can easily see
that Eq.~(\ref{twopoint}) is obtained from Eq.~(\ref{fIF}) when the
limit $m_{I(F)}\rightarrow\infty$ is taken and the relations of form factors
in Eq. (\ref{FG}) are used.

In order to ensure Luke's theorem \cite{Luke} for the three-point function
one has  to ensure that the two-point  sum rule yields a correct
difference between the baryon and quark masses. 
Using the reasonable values $m_b=4.7$ GeV and
$m_c$=1.35 GeV we obtain 0.94 GeV for the mass difference 
when the cutoff and Borel parameters take respectively the values 
$\om_c=1.4$ and $\mu=1 ~ \GeV$  or  $\om_c=1.3$ and $\mu=2 ~ \GeV$. 

\begin{figure} \label{lcrh6}
\leavevmode
\centering
\epsfysize=8.0cm
\epsffile{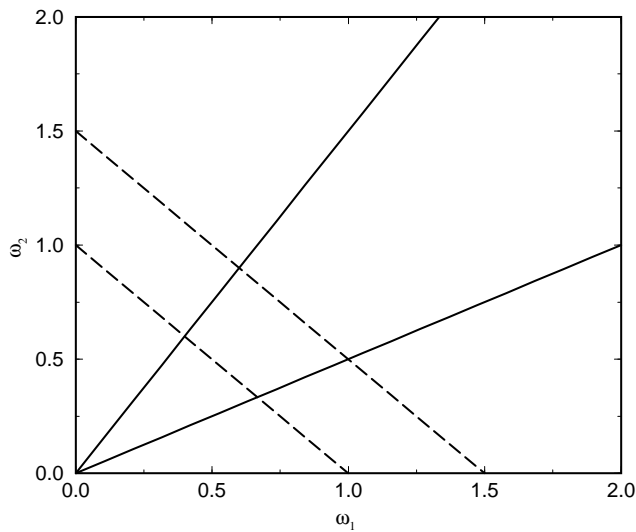}
\caption{Quark-hadron duality in HQET. The 
physical region for the three-point function with heavy hadrons for a
fixed value of y is the wedge between thesolid lines. According to ref.\protect \cite{BloS93} 
duality can only
be assumed after integration over the variable $\om =\om_1-\om_2$, namely 
along the dotted lines. }
\end{figure}

In order to determine the Isgur-Wise (IW) function, the choice of the continuum
model is very important. As has been argued by Neubert \cite{Neu92}
and shown in detail in a non-relativistic model by Blok and Shifman 
\cite{BloS93} it is essential to use a triangular cutoff in order to 
take into account correctly the contributions 
of the higher resonances, since a  rectangular cutoff  yields non-analytic
behavior of the IW function at $y=1$ and, as has been shown at least in the
model consideration in ref. \cite{BloS93}, quark hadron duality
for the three-point holds only after the double spectral function is 
integrated along the dotted lines in Fig. 1. 
Performing this integration along the lines 
$\om = (\om_1+\om_2)/2 ={\rm const} $,
we are left with the sum rule for the three-point function, depending only
on the Borel variable $\mu$ corresponding to $\om$, and the IW function 
is given by
\beqa
\xi(y) &=& \left (\frac{8}{(1+y)^3} I_5(\mu) 
+  \frac{5}{32}
 \langle g_s^2 G^2\rangle I_1(\mu)\,\frac{2}{1+y} [1 +\frac{1}{3}
\frac{y-1}{1+y} ] 
+ \frac{10}{3} \pi^4\ka \langle\bar q q\rangle^2  \right)\nn\\
&\times&\left(I_5(\mu) 
+ \frac{5}{32} \langle g_s^2 G^2\rangle I_1(\mu)
+ \frac{10}{3} \pi^4 \ka\langle\bar q q\rangle^2  \right)^{-1}
+O\left(\frac{1}{m}\right)\; .
\label{ig}
\eeqa

As discussed before, we have chosen the nonrelativistic Borel 
parameters 
for the three-point function to be twice the value of the two-point channel
(here we have assumed the Borel parameters to be the same  in the
initial and final channels); this condition is essential for correct 
normalization.
It can be shown that Eq.~(\ref{f1v}) reduces to Eq.~(\ref{ig}) when the 
limit $m_{I(F)}\rightarrow\infty$ is taken.

\begin{figure} \label{lcrh4}
\leavevmode
\centering
\vskip -1cm
\epsfysize=8.0cm
\epsffile{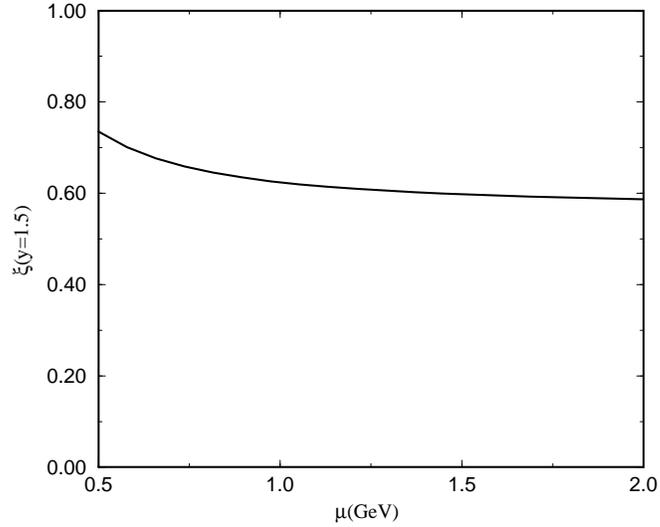}
\caption{Isgur-Wise function at $y$=1.5 as function of the 
Borel parameter $\mu$. }
\end{figure}

\bigskip

\begin{figure} \label{lcrh2}
\leavevmode
\centering
\vskip -2cm
\epsfysize=8.0cm
\epsffile{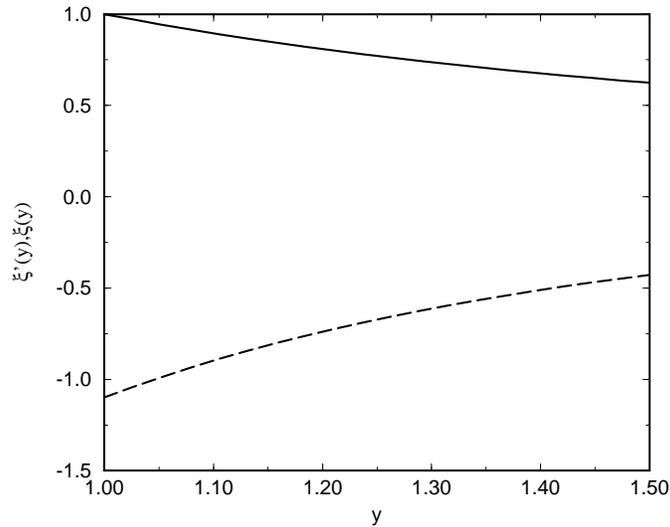}
\caption{IW function (solid line) and its derivative (dashed line) as 
function of $y$ for $\mu=1$ GeV and $\om_c$ = 1.4 GeV. }
\end{figure}

In Fig.~2 we show that  the IW function at $y=1.5$ as a function of
the Borel parameter $\mu$  in the range from 1 to 2 GeV  presents a 
reasonable stability.  In Fig. 3  we display the IW function $\xi(y)$ 
and its slope $\xi'(y)$ in the phenomenologically interesting range for 
$y$  from 1 to 1.5. The influence of the
choice of the parameters can be read off from Table II, where we show  
the values of $\xi(1.5)$ and of the slope $\xi'(1)$. 

\vskip1.0cm
\begin{center}
{\bf Table II}

{\small Values of the IW function $\xi$ at  $y=1.5$  and of its slope
  $\xi'$    \\ 
at $y=1$ for different values of the parameters.}

\vskip0.5cm
\begin{tabular}{c| c| c| c| c}\hline
&$\mu$ &$\om_c$ &$\xi(1.5)$ & $\xi'(1)$\\
& (GeV)&(GeV)   &           &          \\
\hline \hline
standard set& 1   &1.4  &0.638& -1.098  \\
standard set&2  &1.3  & 0.630 &-1.119 \\
standard set&1  &1.6  & 0.592&-1.242  \\
no gluon condensate &1  &1.4  & 0.609& -1.199 \\
no four-quark condensate & 1  &1.4  & 0.560&-1.333   \\
no four-quark condensate & 1  & 1.15  &0.599&-1.202    \\
\hline
\end{tabular}
\end{center}
\vskip1.0cm
We see that the strongest influences come from the choices of the cutoff
$\om_c$ and of the four-quark condensate, and also that the influence of 
the gluon condensate is small.

We can represent the sum rule result for the IW function accurately 
through  a pole fit 
\beq
\xi(y) = \frac{1}{1+1.136(y-1)} ~ .
\eeq

It is important to remark that the slope of the IW function varies appreciably
in the range $1 \leq y \leq 1.5$ as can be seen in the dashed line in Fig. 3.
 Our result for the slope at maximum momentum transfer is about twice the 
value obtained in ref. \cite{Dai96} where a linear fit is used. The slope at
maximum momentum transfer is potentially important for model insensitive 
determinations of $V_{cb}$ \cite{Neu94} from $\La_b$ semileptonic decays.

\begin{figure} \label{lcrh7}
\leavevmode
\centering
\vskip -1cm
\epsfysize=8.0cm
\epsffile{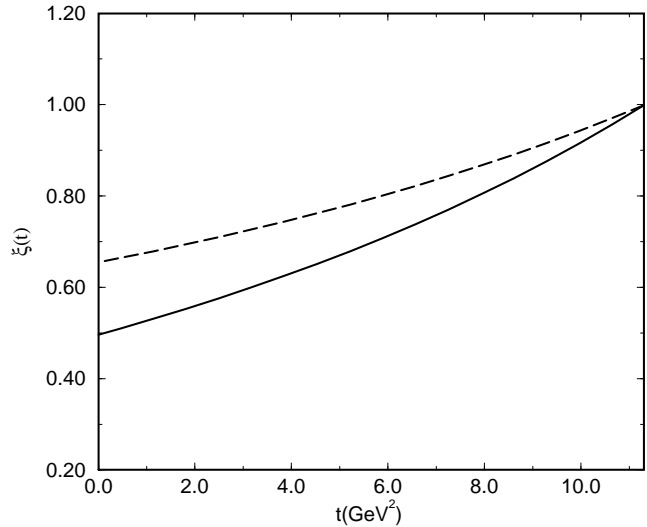}
\caption{ IW function as function of the momentum transfer $t$
(see Eq.~(\protect \ref{t})) (solid line). The dashed line shows the 
same function with $t$ and $y$ related by the physical masses, as given 
in Eq. (\protect \ref{tp}).}
\end{figure}

In order to compare with the results of the full QCD sum rules for
the  $\La_b\to\La_c \ell\nu_\ell$ decay we  plot in Fig. 4  the IW function of Eq. (\ref{ig})
as a function of $t$ (solid line); the dashed line 
shows the IW function as a  function of $t$ 
when  $t$ and $y$ are related with  the physical hadron (and not the quark)  masses through
\beq \label{tp}
t= M_{\La_b}^2+M_{\La_c}^2 - 2 M_{\La_b}M_{\La_c} y ~.
\eeq

\bigskip

{\bf 4. Form Factors and Decay Rates in full QCD}

In this section we discuss the results for the structure $f_If_F$. The
results based on the structures $f'_If_F$ and $f_If'_F$ are given 
respectively in Appendices B and C.

In order to determine the continuum thresholds $u_0$ and $s_0$ for the form factors and decay rates in full QCD we first 
analyse the mass sum rules. The expression for the two point sum rule in
the $\rlap{/}{q}$-structure was  given in Eq.~(\ref{fIF}). The corresponding sum
rule in the $\bf{1}$-structure is given by \cite{bcdn92}
\beqa
f_{I(F)}^2M_{\La_{I(F)}}&=&{e^{M_{\Lambda_{I(F)}}^2/M_{MI(MF)}^2}\over16\pi^4}
\left[{1\over8}m_{I(F)}^5\int_{m_{I(F)}^2}^{s_0(u_0)}ds ~  
    g(s)e^{-s/M_{MI(MF)}^2}
\right.\nn\\
&+&{m_{I(F)}\over96}\langle g_s^2G^2\rangle\int_{m_{I(F)}^2}^{s_0 (u_0)}ds ~
  g_G(s)
e^{-s/M_{MI(MF)}^2}+
\left.{8\over3}\pi^4m_{I(F)}\kappa\langle\overline{q}q\rangle^2
e^{-m_{I(F)}^2/M_{MI(MF)}^2}\right]\; ,
\label{f'IF}
\eeqa
with
\beqa
g(s)&=&(1-x)\left(1+{10\over x}+{1\over x^2}\right)+6\left(1+{1\over x}
\right)\ln{x}\; ,\nn\\
g_G(s)&=&(1-x)\left(7+{2\over x}\right)+6\ln{x}\; ,
\eeqa
where $x=m_{I(F)}^2/s$.

\begin{figure} \label{massb}
\leavevmode
\begin{center}
\vskip -1cm
\epsfysize=8.0cm
\epsffile{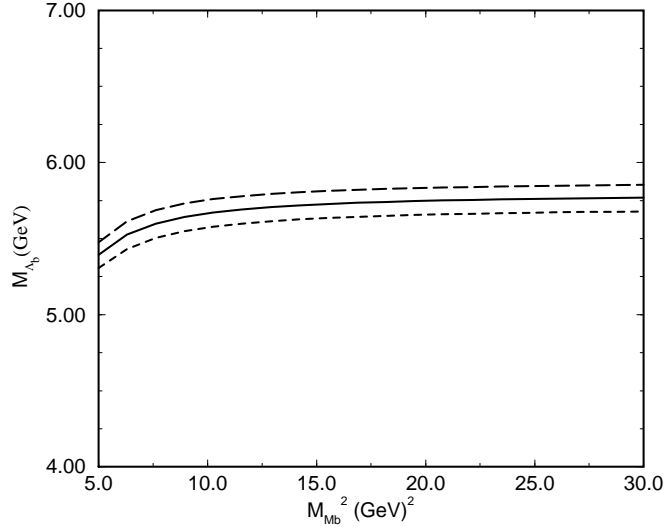}
\caption{$\La_b$ mass obtained from the mass sum rules, as function of the 
Borel mass squared for three
different values of the continuum threshod $s_0=(5.65\GeV+\Delta_s)^2$:
$\Delta_s=0.4\GeV$  (dotted line), $\Delta_s=0.5\GeV$  (solid line) and
$\Delta_s=0.6\GeV$  (dashed line).}
\end{center}
\end{figure}

In Fig. 5 we show $M_{\La_b}$ as a function of the Borel mass $M_{Mb}$,
obtained by dividing Eq.~(\ref{f'IF}) by Eq.~(\ref{fIF}) for different values 
of 
\beq
s_0=(M_{\La_b}+\Delta_s)^2\; .
\label{ds0}
\eeq

The parameter values  used in all calculations are $m_b=4.6$ GeV, 
$m_c=1.4$ GeV, $M_{\La_c}=2.285$ GeV, $M_{\La_b}=5.65$ GeV and 
$\langle\bar{q} q \rangle=-(230 \; {\rm MeV})^3$, and we have neglected 
the gluon condensate. 
As can be seen in this 
figure, there is a rather stable plateau for $M_{Mb}^2\geq10 ~ \GeV^2$, for
the three values of $s_0$ considered. A similar result is obtained 
for $M_{\La_c}$, with a stable plateau for $M_{Mc}^2\geq4 ~ \GeV^2$
as  shown in Fig. 6,  where we plot $M_{\La_c}$ versus the Borel
mass $M_{Mc}$, for three different values of 
\beq
u_0=(M_{\La_c}+\Delta_u)^2\; .
\label{du0}
\eeq

\begin{figure} \label{massck}
\leavevmode
\begin{center}
\vskip -1.cm
\epsfysize=8.0cm
\epsffile{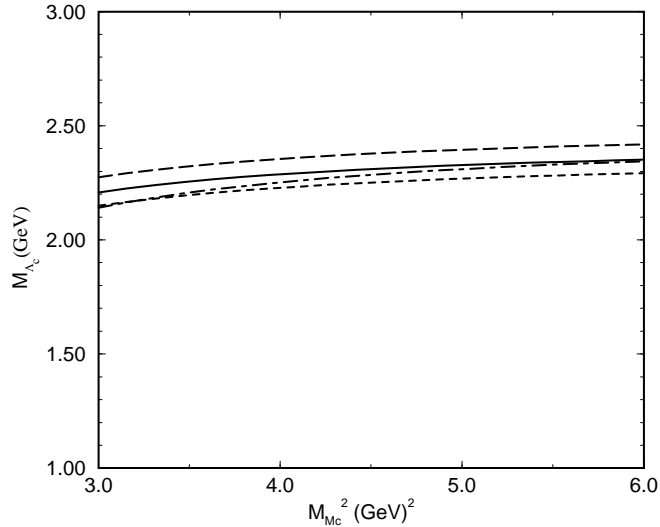}
\caption{$\La_c$ mass obtained from the mass sum rules, as a function of the 
Borel mass squared for three
different values of the continuum threshod $u_0=(2.285\GeV+\Delta_u)^2$:
$\Delta_u=0.6\GeV$  (dotted line), $\Delta_u=0.7\GeV$  (solid line) and
$\Delta_u=0.8\GeV$  (dashed line). The less stable result in 
the dot-dashed line is obtained 
for $\kappa=2$ (see Eq.(\protect \ref{cond})) and $\Delta_u=0.8\GeV$.}
\end{center}
\end{figure}

In Figs. 5 and 6 we have considered $\kappa=1$. A larger value of $\kappa$
decreases the value of $M_{\La_{b(c)}}$ and makes the curve less stable, 
as shown by the dashed-dotted line in Fig. 6, where we have used
$\kappa=2$ and $\Delta_u=0.8$ GeV. 

In ref.~\cite{CDKS82} it was argued that  the
determination of the baryon mass based on the ratio of Eqs.~(\ref{f'IF}) and
(\ref{fIF}) may be misleading especially for light baryons, since states with positive and negative parity contribute to it 
with opposite signs (we shall call this method 1). 
An alternative way to determine the baryon mass is
based on the sum rule Eq.~(\ref{fIF}) and its derivative with respect
to $M_M^{-2}$ that yield  
\beqa
f_{I(F)}^2M_{\La_{I(F)}}^2&=&{e^{M_{\Lambda_{I(F)}}^2/M_{MI(MF)}^2}\over20
\pi^4}\left[
{5\over128}m_{I(F)}^4\int_{m_{I(F)}^2}^{s_0(u_0)}ds~ s~  
     h(s)~ e^{-s/M_{MI(MF)}^2}
\right.\nn\\
&+&
\left.{10\over3}\pi^4\kappa\langle\overline{q}q\rangle^2 m_{I(F)}^2
e^{-m_{I(F)}^2/M_{MI(MF)}^2}\right]\; ,
\label{flIF}
\eeqa
with $h(s)$ given by Eq.~(\ref{hs}). Using the square root of the ratio 
between Eqs.~(\ref{flIF}) and (\ref{fIF}) to determine $M_{\La_{I(F)}}$ 
(method 2) we have obtained curves very similar
to the ones shown in Figs.~5 and 6, but with lower continuum thresholds.
For instance, we cannot distinguish the solid lines in Figs. 5 and 6
from the new curves obtained with $\Delta_s=0.4\GeV$ and $\Delta_u=0.43\GeV$
respectively. Therefore, using method 2 we obtain similar stability as
obtained with method 1, but lower values for the continuum thresholds, this 
effect being even bigger for $\Lambda_c$.

\begin{figure} 
\label{fig7}
\leavevmode
\begin{center}
\vskip -1cm
\epsfysize=8.0cm
\epsffile{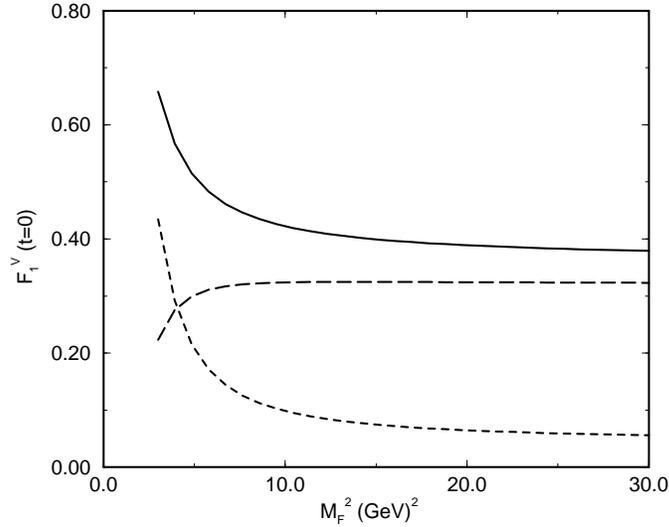}
\caption{Sum rule values for the decay amplitude $F_1^V$ at $t$=0 
for the process $\La_b \rightarrow \La_c \ell \nu_{\ell}$ 
 as function of the Borel mass $M_F^2$. The long-dashed line is the 
perturbative contribution, and the short-dashed line is the four-quark 
condensate for $\kappa=1$ (see Eq. (\protect \ref{cond})). The solid 
line represents the total contribution.}
\end{center}
\end{figure}

We first discuss the process $\La_b\to\La_c \ell\nu_\ell$. In the form 
factor sum rule result, Eq.~(\ref{f1v}), we introduce the relation 
between  the two Borel masses 
\beq
{M_I^2\over M_F^2}={M_{\Lambda_I}^2-m_I^2\over M_{\Lambda_F}^2-m_F^2}\simeq 
3.3\; .\label{ratio}
\eeq

In Fig. 7 we show  the behavior of the contributions to the form factor  
$F_1^V$ at $t$=0 for the process $\La_b \rightarrow \La_c \ell \nu_{\ell}$
for $\kappa=1$  as a function of the Borel mass $M_F^2$ , obtained by
using a rectangular continuun cut-off, Eq.~(\ref{contquad}), in 
Eq.~(\ref{f1v}).
We observe that the range in the Borel masses where the perturbative and 
non-perturbative contributions are in equilibrium is above $10$  GeV$^2$. 
Since the Borel masses in the two- and three-point functions are related
by $M_{MI(MF)}^2=M_{I(F)}^2/2$ and given the relation in Eq.~(\ref{ratio}),
this means that $M_F^2=10 ~ \GeV^2$ is related
with $M_{MF}^2\simeq5 ~ \GeV^2$ and $M_{MI}^2\simeq15 ~ \GeV^2$
which are reasonable values for the Borel parameter in the $\La_c$ 
and $\La_b$ mass sum rules, as can be seen in Figs. 5 and 6. To reproduce the
experimental values of the $\La_c$ and $\La_b$ masses for these values of 
the Borel parameters we need $\Delta_u\simeq0.65$ GeV and
$\Delta_s\simeq0.45 $ GeV (method 1), which are the values used in Fig. 7.
Using method 2 we obtain $\Delta_u\simeq0.45$ GeV and
$\Delta_s\simeq0.35$ GeV which give a similar result for the form factor
resulting from a smaller continuum contribution but a bigger four-quark
condensate contribution.

We have also calculated the $t$-dependence of this form factor in the range 
$0\leq t \leq 8 \; \GeV^2 $ (which covers the major part of the 
kinematically allowed  region  $0\leq t \leq 11.34$ GeV$^2$) where we do 
not have  difficulties caused by non-Landau singularities \cite{BBD91}. The 
$t$-dependence,  represented with dots in Fig. 8,  can be very well 
approximated by a pole fit, as shown by the solid line 
in Fig. 8 (method 1). The extrapolation of the fit to the  
maximal momentum transfer value, $t_{\rm max}=(m_{\La_b}-
m_{\La_c})^2$,  yields $F_1^V = - F^A_1 = 0.97$. 
In HQET this value is just the IW function at zero quark recoil
(see Eqs.~(\ref{FG}))
and is predicted to be $1$. 

\begin{figure}
\label{fig8}
\leavevmode
\begin{center}
\vskip -2cm
\epsfysize=8.0cm
\epsffile{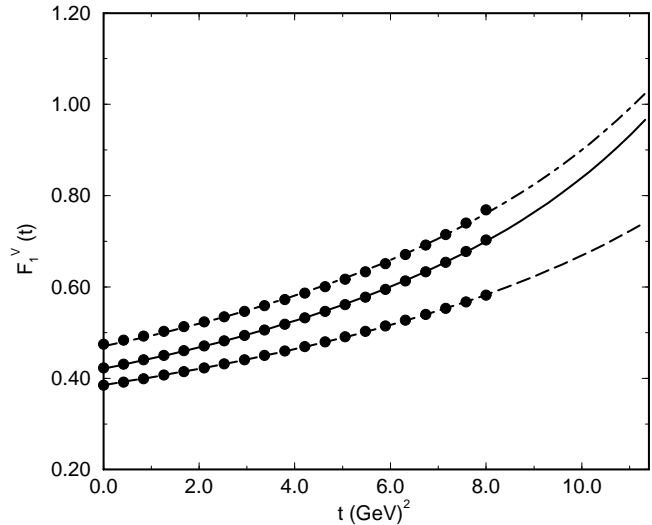}
\caption{ Decay amplitude $F_1^V$ for the process $\La_b \rightarrow 
\La_c \ell \nu_{\ell}$  as function of the squared momentum transferred
to the leptons. Solid line: pole fit $F^V_1(t) =8.46/ (20.08 -t)$ 
to the  sum rule results (dots) for $\kappa=1$ 
(see Eq.~(\protect\ref{cond})) and
$\Delta_s=0.45\GeV$ and $\Delta_u=0.65\GeV$ (see Eqs.~(\protect\ref{ds0},
\ref{du0})) . Dashed line: the same 
for $\Delta_s=\Delta_u=0.5\GeV$; $F^V_1(t) =8.12 /(22.27 -t)$. Dot-dashed
line: the same for $\kappa=2$, $\Delta_s=0.5\GeV$ and $\Delta_u=0.8\GeV$;
$F^V_1(t) =9.84 /(20.93 -t)$.}
\end{center}
\end{figure}

In Fig.~8 we also plot the result obtained
for $F_1^V(t)$ when using $\Delta_s=\Delta_u=0.5\GeV$ and $M_F^2=15\GeV^2$
as done in ref.~\cite{dfnr98} (dashed line). As can be seen in this figure, 
a very different value for $F_1^V(t_{\rm max})$ is obtained in this case. 
The results obtained using $\Delta_s=0.35\GeV$, $\Delta_u=0.45\GeV$ and 
$M_F^2=10\GeV^2$ (method 2) yield $F_1^V(t_{\rm max})=0.84$, with a 
$t$-dependence between the solid and dashed lines in Fig.~8, where 
we also show $F_1^V(t)$ calculated
for $\ka=2$ (dot-dashed line). In this case the continuum thresholds that 
reproduce the experimental value of the $\La_b$ and $\La_c$ masses are 
$\Delta_s=0.5 ~ \GeV^2$ and $\Delta_u=0.8 ~ \GeV^2$ (see Eqs.~(\ref{ds0})
and (\ref{du0})) using method 1, and $\Delta_s=0.45 ~ \GeV^2$ and $\Delta_u=
0.55~ \GeV^2$ (method 2).

Another point already mentioned is the importance of the continuum
model. The above results were obtained using a   rectangular region of 
integration (see Eq.~(\ref{contquad})). However, as pointed out before 
\cite{Neu92,BloS93},
to take correctly into account the contributions of higher resonances 
it is essential, in the HQ limit,  to use a triangular region. Since in 
the present problem the initial and final heavy quarks have different 
masses the triangular region cannot be the same symmetric triangle  
shown in Fig. 1, but it  has to be modified by the ratio $m_F/m_I$. 
Therefore, the limiting 
triangular region is defined by Eq.~(\ref{conttri}). 

\begin{figure} 
\label{fig9}
\leavevmode
\begin{center}
\vskip -1cm
\epsfysize=8.0cm
\epsffile{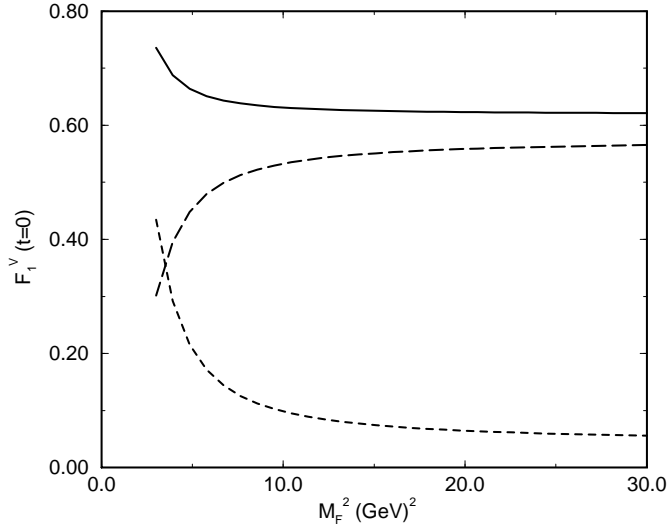}
\caption{Same as Fig.~7 for the triangular region of the continuum
model.}
\end{center}
\end{figure}

In Fig. 9 we show  the behavior of the contributions to the form factor  
$F_1^V$ at $t=0$ for $\kappa=1$  as function of the Borel mass $M_F^2$,
obtained using the above defined triangular region of the continuum
model and method 1 to determine the continuum thresholds.
We observe that the perturbative contribution is now bigger than when
using the rectangular region of integration and that the stability
of the curve, as a function of the Borel parameter, is even better than
in Fig. 7.

The $t$-dependence of $F_1^V(t)$, extracted at $M_F^2=10 ~ \GeV^2$, is 
represented with dots in Fig.~10. 
Again it can be very well approximated by a pole fit as shown by the solid 
line. The dashed line shows the pole fit to the sum
rule result for $\ka=2$. The extrapolation of the fits to the  
maximal momentum transfer value, $t_{\rm max}=(m_{\La_b}-
m_{\La_c})^2$,  yields $F_1^V = 1.05$ for $\ka=1$ and $F_1^V = 1.09$ for 
$\ka=2$ (method 1). These values have to be understood as upper limits 
to $F_1^V(t_{\rm max})$.

\begin{figure}
\label{fig10}
\leavevmode
\begin{center}
\vskip -1cm
\epsfysize=8.0cm
\epsffile{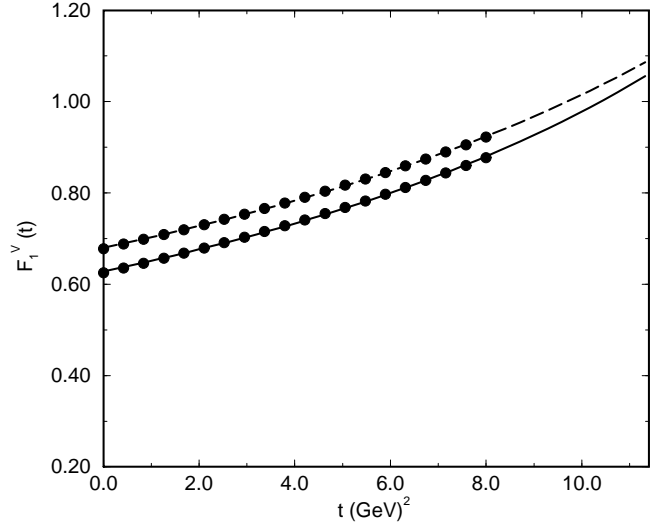}
\caption{Same as Fig.~8 for the triangular region of the continuum
model. Solid line: pole fit $F^V_1(t) =17.55 /(27.94 -t)$ 
to the  sum rule results (dots)
for $\kappa=1$ (see Eqs.~(\protect\ref{cond}) . Dashed line: the same 
for $\kappa=2$; $F^V_1(t) =20.57 /(30.26 -t)$.}
\end{center}
\end{figure}

Using the triangular region for the continuum
model and method 2 to determine the continuum thresholds we get $F_1^V(t_{\rm 
max}) = 1.01$ for $\ka=1$ and $F_1^V(t_{\rm max}) = 0.99$ for 
$\ka=2$. Therefore, we conclude that the results are much less sensitive
to the continuum thresholds when the triangular region of the continuum
model is used.

The slope of $F_1^V$ as a function of $y$ (see Eq.~(\ref{tp}))
at maximum momentum transfer, that gives the slope of the Isgur-Wise function 
at zero recoil, $\xi'(1)$, comes out to be very large for the rectangular 
region,  $\xi'(1) = -2.4 \pm 0.4$, but is more consistent with the
results given in Table II for the triangular region 
$\xi'(1) = -1.45 \pm 0.15$. The errors reflect variations of $\kappa$ from 1 
to 2 and variations on the continuum thresholds obtained with methods 1 and 2.
Also, comparing Figs. 10  and 8 with the dashed line in Fig. 4 we see that the
results from HQET are very similar to the ones obtained with the triangular 
region in full QCD.

In Table III we compare  the HQET values for the form factors 
with  $1/m_Q $ and short distance corrections given in ref.  \cite{n94} 
with our results (see Eq. (\ref{FG})). The indicated errors account for
both choices of models for the continuum integration. 

\vskip1.0cm
\begin{center}
{\bf Table III}

{\small Form-factors for the semi-leptonic decay $\Lambda_b\to \Lambda_c \ell
 \nu_{\ell} $ at maximum momentum transfer for HQET in the limit of
infinite quark masses, including $1/m_Q$ corrections and short distance 
corrections \protect \cite{n94} and our results from QCD sum rules.}
\vskip0.5cm
\begin{tabular}{c|c|c|c } \hline
 &                 &  with  & \\
   form & $m_Q\to\infty $ & $ ~ 1/m_Q$ & ~ our results\\
  factors  &              & corrections &              \\
\hline
$F_1^V$ & 1 & 1.03 & $1.03 \pm 0.06$ \\
$F_2^V$ & 0 & 0.10 &  0 \\
$F_3^V$ & 0 & 0.00 &  0 \\
$F_1^A$ & -1 & -0.97 &  $-1.03 \pm 0.06$ \\
$F_2^A$ & 0 & 0.00 &  0 \\
$F_3^A$ & 0 & 0.12 &  0 \\
\hline
\end{tabular}
\end{center}
\vskip1.0cm

With the pole parametrization of the form factors we can evaluate
the decay rate of the semileptonic decay $\Lambda_b\rightarrow\Lambda_c
\ell\nu_\ell$. Using $V_{cb} = 0.04$, we obtain  for the decay width
\beq
\Gamma(\La_b\to \La_c\ell^-\bar\nu_{\ell}) = (3.3 \pm 1.1) \times 10^{-14} ~  
{\rm GeV} ~,
\label{deb1}
\eeq
where the errors reflect variations of $\kappa$ from 1 to 2, the choices  
of different regions of the continuum contribution and different continuum
thresholds. This value is in agreement with other 
predictions \cite{Kor94,ILKK97,DHHL96,lls98} , which are in the range  
($3.5$ to $6.17$) $\times 10^{-14}$ GeV, and is also in agreement with the 
experimental  upper limit \cite{PDG98} given by 
\beq
\Gamma(\La_b\to \La_c^+ 
\ell^-\bar\nu_{\ell} + {\rm anything})= (4.4 \pm 1.8)\times 10^{-14} 
\;\GeV ~. 
\label{upb}
\eeq
There is only one calculation of the semileptonic decay $\Lambda_b
\rightarrow\Lambda_c\ell\nu_\ell$ from lattice QCD \cite{ni97}, which 
gives, however, only a lower limit to the decay rate since the method 
cannot access all the velocity transfer range.

\begin{figure} \label{fig11}
\leavevmode
\begin{center}
\vskip -1cm
\epsfysize=8.0cm
\epsffile{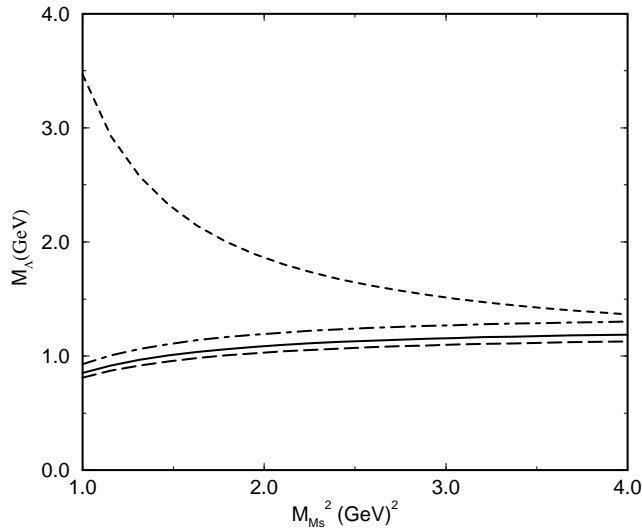}
\caption{$\La$ mass obtained from the mass sum rules, as a function of the 
Borel mass squared for 
different values of the continuum threshod $u_0=(1.115\GeV+\Delta_u)^2$.
The dotted line shows the result obtained from
Eqs.~(\protect \ref{fIF}) and (\protect \ref{f'IF}) and $\Delta_u=0.5\GeV$.
The other lines show the results obtained from
Eqs.~(\protect \ref{fIF}) and (\protect \ref{flIF}) and
$\Delta_u=0.45\GeV$  (dashed line), 
$\Delta_u=0.5\GeV$  (solid line) and
$\Delta_u=0.6\GeV$  (dot-dashed line).}
\end{center}
\end{figure}

We now turn to the analysis of the $\Lambda_c\to\Lambda\ell\nu_{\ell}$
semileptonic decay. We first analyse the $\La$ mass sum rule that can be 
obtained from Eqs.~(\ref{fIF}), (\ref{f'IF}) and  (\ref{flIF}) considered 
only up to order ${\cal O}$$(m_s^2)$.
In Fig. 11 we show $M_{\La}$ as a function of the Borel mass $M_{Ms}$,
obtained by dividing Eq.~(\ref{f'IF}) by Eq.~(\ref{fIF}) (method 1) for 
$\Delta_u=0.5 \GeV$ (dotted line), where now
\beq
u_0=(M_{\La}+\Delta_u)^2\; ,
\label{du0s}
\eeq
and $m_s=0.17$ GeV, $M_{\La}=1.115$ GeV. As can be seen from  the 
dotted line in this figure,
the $\La$ mass obtained by method 1 is not stable as a function
of the Borel mass. In Fig. 11 we also show $M_{\La}$ as a function of the 
Borel mass,
obtained by dividing Eq.~(\ref{flIF}) by Eq.~(\ref{fIF}) (method 2)
for different values
of $\Delta_u$. With this procedure we obtain a very stable plateau 
for $M_{Ms}^2\geq 1.5 ~ \GeV^2$. 

Considering the masses of the baryons in the initial and final states
the relation between the Borel masses in the three-point function is now 
given by
\beq
{M_I^2\over M_F^2}={M_{\Lambda_I}^2-m_I^2\over M_{\Lambda_F}^2-m_F^2}\simeq 
2.8\; .\label{ratio2}
\eeq

\begin{figure} 
\label{fig12}
\leavevmode
\begin{center}
\vskip -1cm
\epsfysize=8.0cm
\epsffile{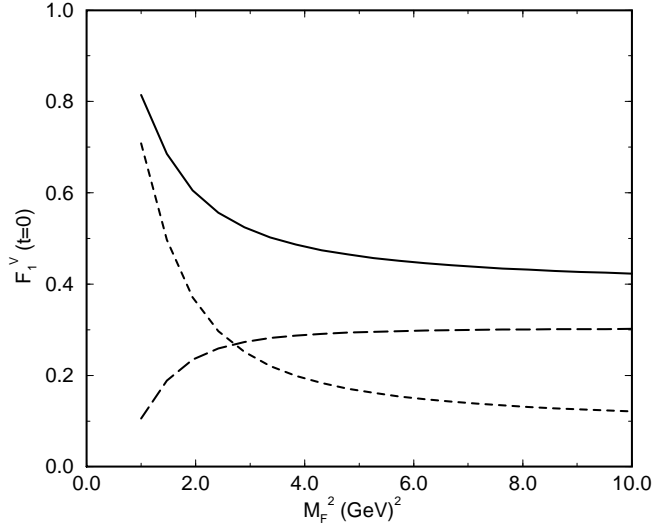}
\caption{The sum rule values for the decay amplitude $F_1^V$ at $t$=0 
for the process $\La_c \rightarrow \La \ell \nu_{\ell}$ 
 as function of the Borel mass $M_F^2$ for $\kappa=1$. The long-dashed 
line shows the perturbative contribution, and the short-dashed line is 
the four-quark condensate for $\kappa=1$ (see Eq. (\protect \ref{cond})). 
The solid line shows  the total contribution.}
\end{center}
\end{figure}

In Fig. 12 we show  the behavior of the contributions to the form factor  
$F_1^V$ at $t$=0 for the process $\La_c \rightarrow \La \ell \nu_{\ell}$
for $\kappa=1$  as function of the Borel mass $M_F^2$ .
We observe that the region in the Borel mass where the perturbative and 
non-perturbative contributions are in equilibrium is above $4$  GeV$^2$. 
Since the Borel masses in the two- and three-point functions are related
by $M_{MI(MF)}^2=M_{I(F)}^2/2$,  and given  Eq.~(\ref{ratio2}),
we have  that $M_F^2=4 ~ \GeV^2$ is related
with $M_{MF}^2\simeq2 ~ \GeV^2$ and $M_{MI}^2\simeq5.5 ~ \GeV^2$. These
are reasonable values for the Borel parameter in the $\La_c$ 
and $\La$ mass sum rules, as can be seen in  Figs.~6 and 11. To reproduce the
experimental values of the $\La_c$ and $\La$ masses for these values of 
the Borel parameters we need $\Delta_u\simeq0.55 ~ \GeV$ and
$\Delta_s\simeq0.62 ~ \GeV$ (method 1) , which are the values that define the
rectangular region of integration used in Fig. 12. Comparing Figs.~7 and 12,
we see that the importance of the four-quark condensate is much 
larger  in  $\La_c$ than in $\La_b$ decay. Using method 2 to fix the 
continuum threshold for $\Lambda_c$ we get $\Delta_s\simeq0.42 ~ \GeV$
and, in this case, the importance of the four-quark condensate is even bigger,
yielding a form factor with almost the same Borel mass dependence as obtained
with nethod 1 but about $1.2$ times bigger.

\begin{figure}
\label{fig13}
\leavevmode
\begin{center}
\vskip -1cm
\epsfysize=8.0cm
\epsffile{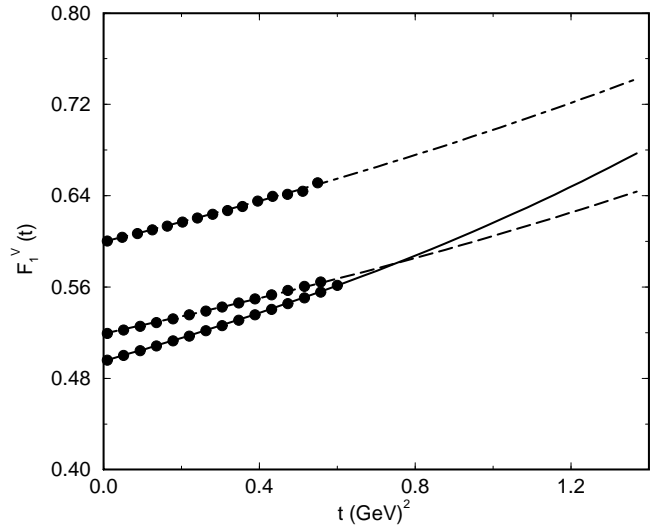}
\caption{The decay amplitude $F_1^V$ for the process $\La_c \rightarrow 
\La \ell \nu_{\ell}$  as function of the squared momentum  $t$ 
transferred to the leptons. Solid line gives the pole fit 
$F^V_1(t) =2.52 /(5.09 -t)$ for $\kappa=1$ and the
rectangular region of the continuum contribution. The dashed line represents
 the same for $\kappa=2$, with  $F^V_1(t) =3.67 /(7.07 -t)$. The dot-dashed
line shows the same for $\kappa=1$ and the triangular region of the continuum 
contribution, with  $F^V_1(t) =4.25 /(7.09 -t)$.}
\end{center}
\end{figure}

The $t$-dependence of the form factor $F_1^V$  in the range 
$0\leq t \leq 0.6 \; \GeV^2 $ where we   do not 
find  difficulties with non-Landau singularities \cite{BBD91} is shown
by the dots in Fig. 13 for two different choices 
of the four-quark condensate ($\kappa=1, 2$ ). The pole fits (solid line 
for $\ka=1$ and dashed line for $\ka=2$) are also shown in this figure. 
For $\ka=2$ the continuum thresholds that reproduce the experimental
masses for $M_{Ms}^2\simeq2 ~ \GeV^2$ and $M_{Mc}^2\simeq5.5 ~ \GeV^2$ are
respectively $\Delta_u\simeq0.65$ GeV and $\Delta_s\simeq0.8 $ GeV (method 1)
and $\Delta_s\simeq0.52 $ GeV (method 2). These differences in the continuum 
thresholds lead, in this case of the  $\La_c$ decay, to  very
different behavior of the form factors as a function of $t$. This change
is not observed in the $\La_b$ decay.

If  we use the triangular region instead of the rectangular region for 
the continuum we obtain larger  
values for the form factor, as shown by the dot-dashed line in Fig. 13
for $\ka=1$ and $\Delta_s\simeq0.62 $ GeV (method 1). In this case we can 
see that the pole file is not as perfect
as in the other cases. For $\ka=2$ and the triangular region the pole
fit to the sum rule results is given by $F_1^V(t)=5.34/(8.85-t)$ (method 1).

Using the form factors obtained with method 1 for the determination of the
continuum threshold for $\Lambda_c$ and
$V_{cs} = 0.975$, we get for the decay width
\beq
\Gamma(\La_c^+\to \La + e^+ +  \nu_e ) = (8.7  \pm 1.2) \times 10^{-14} \;
{\rm GeV}. 
\label{dec1}
\eeq
The overwhelming sources of the theoretical error are the uncertainties 
in the value of the four-quark condensate and the choice of the model 
for the continuum.
Variations of the quark masses within reasonable limits such as  
$0.13\leq m_s \leq0.17 ~ \GeV$ and $1.25\leq m_c\leq 1.45 ~ \GeV$ have  
negligible effects compared to the errors introduced by the variation of 
the four-quark condensate and the model of continuum.
Within the errors the above result agrees with the reported experimental 
value\cite{PDG98}
\beq
\Gamma_{\rm exp}(\La_c^+\to \La + e^+ +  \nu_e ) = (6.3  \pm 1.9) 
\times 10^{-14} \; 
{\rm GeV} ~ .
\eeq

The form factors obtained with method 2 for the determination of the
continuum threshold for $\Lambda_c$ yield a very large decay width
($(13.6\pm 2.3) \times 10^{-14} \;{\rm GeV}$) but we do not think we can trust
this result since, in this case, the sum rule is dominated by the four-quark 
condensate.

Several theoretical attempts have been made to describe  the form
factors in the  $\La_c^+\to \La + e^+ +  \nu_e$ semileptonic decay,
employing flavor symmetry and/or quark models \cite{ga79,pm89,sin91,ct96}.
Their results for the decay width are in the range  
$(4.7$ - $10$) $\times 10^{-14}$ GeV and are thus in agreement with 
our result in Eq.~(\ref{dec1}). In spite of this good agreement in 
the decay width, Eq.~(\ref{general}) leads to an asymmetry parameter 
$\alpha = -1$ while  the observed value \cite{PDG98} is  
$\alpha =-0.82\pm 0.10$.

We can also make predictions about the decay width of the semileptonic decay
$\Lambda_b\to p\ell\nu_{\ell}$. This decay was recently studied in the partial
HQET framework \cite{hep98} with result $\Gamma(\Lambda_b\to p\ell
\nu_{\ell})=(1.43\pm0.07)\times10^{-11}|V_{ub}|^2\GeV$. Since the proton
is not a heavy baryon, it is very important to compare 
this result obtained in the framework of partial HQET with our full QCD result 
though it should be remembered that there are serious objections against the 
use of conventional sum rules for heavy light transitions 
\cite{bb97}. 

The
procedure is exactly the same as described above for the $\La_c$ semileptonic 
decay. The results for the proton mass and form factor as functions of the
Borel mass are very similar to those presented in Figs. 11 and 12,  being
only a little less stable. The region in the Borel mass where the 
perturbative and non-perturbative contributions to the form factor are in 
equilibrium is above $3$  GeV$^2$. Therefore, we evaluate the sum
rules at $M_F^2=3 ~ \GeV^2$ which give $M_{Mp}^2\simeq1.5 ~\GeV^2$ and 
$M_{Mb}^2 \simeq18 ~ \GeV^2$ for the value of the Borel parameters in the 
mass sum rules
of the proton and $\La_b$ respectively. These seem to be rather
large values for the Borel parameters in the two-point function, but 
we found that there is no stability in the form factor for smaller values 
of these parameters.

To reproduce the
experimental values of the proton and $\La_b$ masses for these values of 
the Borel parameters we need $\Delta_u\simeq0.6\GeV$  and
$\Delta_s\simeq0.4\GeV$  for $\ka=1$ and 
$\Delta_u\simeq0.75\GeV$ and $\Delta_s\simeq0.5\GeV$ for $\ka=2$, where
$\Delta_u$ and $\Delta_s$ were determined by method 2 and 1 respectively.
Using also either the
rectangular  or the triangular region  for the continuum
contribution we obtain
\beq
\Gamma( \La_b\to p + \ell^- + \bar\nu_{\ell} )=
(1.7\pm0.7)\times10^{-11}|V_{ub}|^2\GeV ~  ,
\eeq
in agreement with the result of the HQET calculation\cite{hep98}.
 
\begin{figure}
\label{fig14}
\leavevmode
\begin{center}
\vskip -1cm
\epsfysize=8.0cm
\epsffile{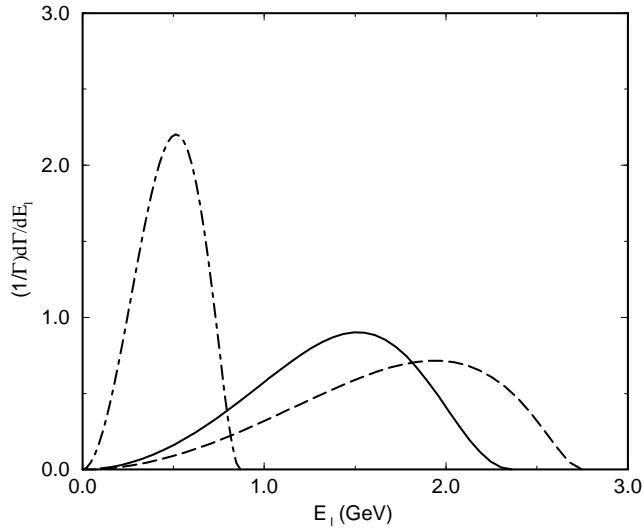}
\caption{Normalized charged lepton energy spectrum
for the processes  $\La_c \rightarrow \La \ell \nu_{\ell}$ (dot-dashed line),
$\La_b \rightarrow \La_c \ell \nu_{\ell}$ (solid line)  and
$\La_b \rightarrow p \ell \nu_{\ell}$ (dashed line).}
\end{center}
\end{figure}

The knowledge of the form factors also allows us to calculate the charged 
lepton  energy spectrum. In Fig. 14 we plot $(1/\Gamma){d \Gamma}/{dE_l}$
versus the energy of the charged lepton, $E_l$, for the three studied decays.
The spectra have exactly the same shapes for $\kappa=1$ or $2$ and for 
either the rectangular or triangular region in the continuum model.

\bigskip

{\bf 5. Summary and Conclusions }

We have developed in this paper the complete kinematical formalism  
to obtain  the physical form factors of semileptonic hyperon  
decays from the off-shell three-point functions and QCD sum rules. 
We have obtained  the $t$ 
-dependence of the form factors  directly 
from the sum rules in the physical region for positive values of the 
momentum transfer since, in this case, the cut in the $t$ channel starts 
at $t\sim m_Q^2$ ($m_Q$ being the heavy quark mass), and thus the 
Euclidean region stretches up to that threshold.
We have applied this formalism to the 
$\La_c^+\to \La + e^+ +  \nu_e$, 
$ \La_b\to \La_c + \ell^- + \bar\nu_{\ell}$ 
and  $ \La_b\to p + \ell^- + \bar\nu_{\ell}$  semileptonic decays.
It turns out that the sum rule for 
$\La_c^+\to \La + e^+ +  \nu_e$
depends much more strongly on the values of parameters used in the 
sum rules than what has been found in the case of the analogous 
mesonic decay $D\to K^{(*)} + \mbox{ leptons }$. In particular, the 
dependence on the choice of the model for the continuum makes detailed 
predictions impossible. This is  not surprising since:  
 a) the higher dimension of the baryon currents implies a faster increase 
of the discontinuities along the cuts and thus a stronger dependence on the 
assumed onset of duality (continuum threshold); and  b) spin leads to 
further complications. 
In our projection we have only singled out the ground state with spin 
1/2 and put all remaining contributions in the pertubatively treated 
continuum. For higher spin states, which are also present in the 
three-point function, the projection is much more involved and such a 
simple treatment may be less justifiable than in the scalar and 
vector meson cases. This difficulty reflects itself in a certain 
inconsistency of the sum rules for the same form factor obtained from 
different gamma-matrix structures (see Sec. IV and Appendices B and C). 
We have therefore studied in more detail the $f_If_F$ structure 
(see Eq.~(\ref{f1v})). 
Another special feature of the $\Lambda_c$ semileptonic decay  
is the importance of the four-quark condensate which, for some values of
the continuum thresholds, gives much larger contribution 
than the perturbative part. According to the original SVZ philosophy 
\cite{SVZ79}, this brings doubt to the use of the sum rule 
approach in this case,and, furthermore, we must remark that the
value of the four-quark condensate is only poorly known.
So it is natural that the present  analysis of the $\Lambda_c$ decay 
leads to very large errors, and the same is true, to an even larger  
extent, for  $ \La_b\to p \ell^-\bar\nu_{\ell}$.

The situation  for  $ \La_b\to \La_c\ell^-\bar\nu_{\ell}$ is much more 
favorable. Here the symmetries valid in the limit of the masses going 
to infinity make themselves remarkable also in our treatment of full QCD. 
The result of the present more complete investigation modifies  
our earlier assertion \cite{dfnr98} about  large $1/M^2$ 
corrections for this decay. This is mainly due to the choice of the 
appropriate continuum model. It has been strongly argued by Blok and 
Shifman \cite{BloS93}
that only a triangular cutoff  (see Sec. IV) is adequate for 
HQET, since a rectangular cutoff does not project out the p-waves properly. 
This argument becomes  much more important for fermions than for 
mesons, since the change from a rectangular to a triangular cutoff, which
 had no strong effect on the decay rate of the mesons, 
 is very important for baryons. So we conclude that a treatment of the
$ \La_b\to \La_c\ell^-\bar\nu_{\ell}$ 
through HQET sum rules is satisfactory. Nevertheless, we must remark that 
the influence of the continumm threshold on the 
phenomenologically important slope of the Isgur-Wise function is not 
negligible also in the HQET limit.

We have calculated the semileptonic decay form factors using sum rules 
with three different  tensor structures. The most robust result that we 
have obtained for the form factors is that 
$F_1^A(t)=-F_1^V(t)$ for all decays and structures considered. 
Different structures lead to  different relations between
$F_2^V$ and $F_2^A$. The only possible way to conciliate the three results
would be to have $F_2^V=F_2^A=0$, which however is a result valid only for 
the $f_If_F$ structure. In the case of the $\La_b$ decay, the numerical 
values obtained for $F_2^{V(A)}$ in the  $f'_If_F$ and $f_if'_F$ structures 
are very small and, therefore, consistent with zero. However, in the case 
of the $\La_c$ decay, the numerical values obtained for $F_2^{V(A)}$ in the 
$f'_If_F$ and $f_I f'_F$ cases are not consistent with zero, and this leads 
to huge variations in the decay rates obtained in  the different cases.

For the process $\Lambda_b\to\Lambda_c\ell\nu_{\ell}$ our results for 
the decay width are  in agreement with
the experimental upper limit, and the extrapolation of the form 
factors to the maximal momentum transfer  agrees very well with the HQET 
prediction. 

We have also calculated the charged lepton energy spectrum and it is  
interesting to remark that the form of the spectrum is not sensitive to 
the structure used, neither to the model of continuum,  neither to the 
parametrization of the four-quark condensate.

\acknowledgements

We would like to thank R. Rosenfeld for collaboration in the early stages
of this work. This work has been supported in part by FAPERJ, FAPESP and 
CNPq (Brazil)  and by DAAD (Germany).

\bigskip

{\bf Appendix 1 . Projecting out the Form Factors  }

\bigskip
Defining
\begin{eqnarray}
\label{factors}
F^{V,A}_i(s,u,t)= \frac {F_i^{V,A}(t)}{(u-M_{\La_F}^2)(s-M_{\La_I}^2)}
~,~({\rm with}~  i=1,2,3) ~, 
\end{eqnarray}
and using 
$$\lambda(s,t,u)=s^2+u^2+t^2-2su-2st-2ut ~, $$
the 24 unknows defined in Eq.~(\ref{unk}) are given in term of the
traces by

\bigskip

\begin{eqnarray} \lb{pro1}
&& \hspace{-7mm} F_1^V(s,u,t) f_I f_F = \frac{1}{4[\la(s,t,u)]^2} \cdot \nn \\
&& \bigg[ t \lambda (s,u,t)~ x_3 +(2s^2+2st-t^2-4su+2tu+2u^2)~ x_4 \nn \\
&&-3\,t\,\left(s-u\right)~ x_5-3t\left(s-u\right)~ x_6+3t^2~ x_7 \bigg] \\
&& \hspace{-7mm} F_1^V(s,u,t) f_I f'_F = \frac{1}{8 [\la(s,t,u)]^2} \cdot \nn \\ 
&& \bigg[ 2(s-u)\lambda(s,t,u)~ x_1-2t\lambda(s,t,u)~ x_2-t\lambda(s,t,u)~ x_8 
   + t\lambda (s,t,u) ~ x_9   \nn \\
&&  -\left( 2s^2-st-t^2-4su+5tu+2u^2 \right)~ x_{10}
 + 3\,t\,\left(s-t-u \right)~  x_{11} \bigg] \\
&& \hspace{-7mm} F_1^V(s,u,t) f'_I f_F = \frac{1}{8 [\la(s,t,u)]^2} \cdot \nn \\ 
&& \bigg[ 2(s-u)\lambda(s,t,u)~ x_1-2t \lambda(s,t,u) ~ x_2
     -t \lambda(s,t,u) ~ x_8 -t \lambda(s,t,u) ~ x_9 \nn \\
&& + \left(2\,{s^2}+5\,s\,t-{t^2}-4\,s\,u-t\,u+ 2u^2 \right)~ x_{10} 
 -3\,t\,\left( s + t - u \right)~  x_{11} \bigg] \\ 
&& \hspace{-7mm} F_1^V(s,u,t) f'_I f'_F = \frac{1}{8 [\la(s,t,u)]^2} \cdot \nn \\ 
&& \bigg[ -t(s-t+u) \lambda(s,t,u) ~ x_3 
  -(s-t+u)~ (2s^2+2st-t^2-4su+2tu+2u^2)~ x_4 \nn \\
 && + 3\,t\,\left( s- t+ u \right) \,\left(s- u \right)~ x_5  
 +  \left( s - u \right) \,\left(2\,{s^2} - s\,t - {t^2} - 4\,s\,u - 
     t\,u + 2\,{u^2} \right) ~ x_6   \nn \\
 && -t\left(2\,{s^2}-s\,t-{t^2}-4\,s\,u-t\,u+2\,{u^2} \right)~ { x_7} 
+ i t \lambda (s,t,u) ~ x_{12}\bigg] \\
&& \hspace{-7mm} F_2^V(s,u,t) f_I f_F = -\frac{1}{4[\la(s,t,u)]^2}\cdot \nn \\ && 
\bigg[ \lambda(s,u,t)~ x_8-3\,\left( s - u \right) ~  x_{10}+3t~ x_{11}\bigg] \\
&& \hspace{-7mm} F_2^V(s,u,t) f_I f'_F = -\frac{1}{8 [\la(s,t,u)]^2} \cdot \nn \\ 
&& \bigg[ (s-t-u) \lambda(s,t,u) ~ x_3
 + \left(4s^2-5st+t^2+4su+tu-8u^2\right)~ x_4 \nn \\ 
&& - \left( {s^2}+ s\,t- 2\,{t^2}- 2\,s\,u + 7\,t\,u+ {u^2}
 \right) ~ x_5  
 - 3\,\left( s - t - u \right) \left( s- u \right)~ x_6  \nn \\
&& + 3\,t\,\left( s - t - u \right)~ x_7-i \lambda (s,t,u) ~ x_{12}\bigg]\\
&& \hspace{-7mm} F_2^V(s,u,t) f'_I f_F = -\frac{1}{8 [\la(s,t,u)]^2} \cdot \nn \\ 
&& \bigg[ -(s+t-u) \lambda(s,t,u) ~ x_3
 -\left(8s^2-st -t^2- 4su+5tu-4u^2 \right)~  x_4 \nn \\
 && + \left( {s^2} + 7\,s\,t - 2\,{t^2} - 2\,s\,u + t\,u + {u^2}
      \right)~ x_5 + 3\,\left(s-u \right) \left(s+t-u \right) x_6 \nn \\
 && - 3\,t\,\left(s+t-u \right)~ x_7 - i \lambda (s,t,u) ~ x_{12} \bigg] \\
&& \hspace{-7mm} F_2^V(s,u,t) f'_I f'_F = -\frac{1}{8 [\la(s,t,u)]^2} \cdot \nn \\ 
&& \bigg[ - 2 \left(s- u \right) \lambda (s,t,u)~ x_1 
+  2 t \lambda (s,t,u) ~ x_2
 + \left( s + u \right) \lambda (s,t,u) ~ x_8 \nn \\ 
&&  - \left(s- u \right)\lambda (s,t,u) ~ x_9  
      +3\,\left(s- u \right) \,\left( s - t + u \right)~  x_{10}  \nn \\ 
&& -\left({s^2}+s\,t-2\,{t^2}-2\,s\,u+t\,u+u^2 \right) ~ x_{11} \bigg] \\
&& \hspace{-7mm} F_3^V(s,u,t) f_I f_F = \frac{1}{4 [\la(s,t,u)]^2} \cdot \nn \\ 
&& \bigg[\lambda (s,u,t)~ x_9 -3\,\left( 2\,s - t + 2\,u \right)~  x_{10}
 + 3\,\left( s - u \right) ~  x_{11} \bigg]\\
&& \hspace{-7mm} F_3^V(s,u,t) f_I f'_F = \frac{1}{8 [\la(s,t,u)]^2} \cdot \nn \\ 
&& \bigg[(s-t+3u) \la(s,t,u)~ x_3 + 3(2s-t+2u) (s-t+3u)~ x_4 \nn \\
&& - 3\,\left( s - u \right) \,\left( s - t + 3\,u \right)~  x_5
-\left(5\,{s^2}-7\,s\,t+2\,{t^2}+2\,s\,u-t\,u-7\,{u^2} \right)~ x_6 \nn \\ 
&& +\left( 2\,{s^2} - s\,t-{t^2}-4\,s\,u + 5\,t\,u + 2\,{u^2}\right)~ x_7 
    - i ~\lambda(s,t,u)~ x_{12} \bigg] \\
&& \hspace{-7mm} F_3^V(s,u,t) f'_I f_F = \frac{1}{8 [\la(s,t,u)]^2} \cdot \nn \\ 
&& \bigg[-(3s-t+u)\la(s,t,u)~ x_3 -3(3s-t+u)(2s-t+2u)~ x_4 \nn \\
&& + 3\,\left( 3\,s - t + u \right) \,\left(s- u \right)~ x_5
+\left(7\,{s^2}+s\,t-2\,{t^2}-2\,s\,u+7\,t\,u-5\,{u^2} \right)~ x_6 \nn \\
&& -\left(2\,{s^2}+5\,s\,t-{t^2}-4\,s\,u-t\,u+2u^2\right)~  x_7 
 + i~ \lambda (s,t,u)~ x_{12} \bigg] \\
&& \hspace{-7mm} F_3^V(s,u,t) f'_I f'_F = \frac{1}{8 [\la(s,t,u)]^2} \cdot \nn \\ 
&& \bigg[-2(2s-t+2u) \lambda(s,t,u)~  x_1 + 2(s-u) \lambda(s,t,u) ~x_2 \nn \\
&& +(s-u) \lambda(s,t,u)~ x_8 - (s+u) \lambda(s,t,u)~ x_9 \nn \\
&&+\left(5\,{s^2}-7\,s\,t+2t^2+ 14su-7tu+5u^2\right) ~ x_{10} 
 -3\left(s-t+u \right) \left(s- u \right)~  x_{11} \bigg] 
\eeqa
and for the axial form factors we obtain
\beqa
&& \hspace{-7mm} F_1^A(s,u,t) f_I f_F = \frac{1}{4 [\la(s,t,u)]^2} \cdot \nn \\ 
&& \bigg[t \lambda(s,t,u) ~ u_3 +\left(2s^2+2st-t^2-4su+2tu+2u^2\right) ~ u_4 \nn \\
&& -3t \left( s - u \right)~ u_5
  - 3\,t\,\left(s-u \right)~ u_6 + 3t^2 ~ u_7 \bigg]  \\
&& \hspace{-7mm} F_1^A(s,u,t) f_I f'_F = \frac{1}{8[\la(s,t,u)]^2} \cdot \nn \\ 
&& \bigg[-2(s-u) \lambda(s,t,u) ~ u_1 + 2 t \lambda(s,t,u)~ u_2
+ t \lambda(s,t,u) ~ u_8 -t \lambda(s,t,u) ~ u_9 \nn \\ 
&& +\left(2s^2-st-t^2-4su+5tu+ 2u^2 \right)~ u_{10}
     - 3\,t\,\left(s - t- u \right)~ u_{11} \bigg] \\
&& \hspace{-7mm} F_1^A(s,u,t) f'_I f_F = \frac{1}{8[\la(s,t,u)]^2} \cdot \nn \\ 
&& \bigg[ 2\,\left( s- u \right) \lambda (s,t,u)~ u_1
 - 2t \lambda (s,t,u)~ u_2 - t \lambda (s,t,u) ~ u_8 
   - t \lambda (s,t,u) ~ u_9 \nn \\
&& +\left(2s^2+5st-t^2-4su-tu+2u^2 \right)~  u_{10} 
       - 3\,t\,\left( s + t - u \right) ~  u_{11}  \bigg] \\
&& \hspace{-7mm} F_1^A(s,u,t) f'_I f'_F = \frac{1}{8 [\la(s,t,u)]^2} \cdot \nn \\ 
&& \bigg[ t(s-t+u) \lambda(s,t,u) ~ u_3
+(s-t+u)(2s^2+2st-t^2-4su+2tu+2u^2) ~ u_4 \nn \\
&& -  3\,t\,\left( s- u \right) \,\left( s-t+u \right) \,{ u_5} 
   - \left( s - u \right) \left( 2\,{s^2}- s\,t - {t^2} - 4\,s\,u 
   - t\,u + 2\,{u^2} \right) \,{ u_6} \nn \\ 
&& +t \left(2s^2-st-t^2-4su-tu + 2u^2 \right)~ u_7 
  - i\,t \lambda(s,t,u) ~ u_{12} \bigg]  \\
&& \hspace{-7mm} F_2^A(s,u,t) f_I f_F = -\frac{1}{4[\la(s,t,u)]^2} \cdot \nn \\ 
&& \bigg[ -\lambda(s,t,u) ~ u_8
 + 3\,\left( s - u \right)~ u_{10} - 3t~   u_{11} \bigg] \\ 
&& \hspace{-7mm} F_2^A(s,u,t) f'_I f_F = -\frac{1}{8[\la(s,t,u)]^2} \cdot \nn \\ 
&& \bigg[ (s-t-u) \lambda(s,t,u) ~ u_3
 +\left(4s^2-5st+t^2+4su+tu-8u^2\right)~  u_4  \nn  \\  
&&  - \left( {s^2} + s\,t - 2\,{t^2} - 2\,s\,u + 7\,t\,u + {u^2}\right)~ u_5 
    - 3\,\left(s- u \right) \,\left(s- t - u \right)~  u_6 \nn \\ 
&&+ 3t\left(s- t- u \right)~  u_7 - i \lambda (s,u,t) ~ u_{12} \bigg] \\
&& \hspace{-7mm} F_2^A(s,u,t) f'_I f_F = -\frac{1}{8 [\la(s,t,u)]^2} \cdot \nn \\ 
&& \bigg[ (s+t-u) \lambda(s,t,u) ~ u_3
  +\left(8s^2-st-t^2-4su+5tu-4u^2\right)~  u_4 \nn \\
&& - \left( {s^2} + 7\,s\,t - 2t^2 - 2su + tu + u^2\right) ~ u_5
 - 3\,\left( s - u \right) \left( s + t - u \right) ~ u_6  \nn \\ 
&& + 3\,t\,\left( s+t- u \right) ~  u_7 
 + i \lambda (s,u,t) ~ u_{12} \bigg] \\
&& \hspace{-7mm} F_2^A(s,u,t) f'_I f'_F = -\frac{1}{8 [\la(s,t,u)]^2} \cdot \nn \\ 
&& \bigg[ - 2\,\left( s - u \right) \lambda (s,u,t) ~ u_1  
 +2 t\lambda (s,t,u) ~ u_2 + \left(s + u \right) \lambda(s,t,u)~ u_8 \nn \\
&&  - \left(s -u \right) \lambda (s,t,u) ~ u_9  
 +3\left(s- u \right)\left(s-t+u \right)~  u_{10} \nn \\
&& -\left(s^2+st-2t^2-2su+tu+u^2 \right)~ u_{11} \bigg]\\
&& \hspace{-7mm} F_3^A(s,u,t) f_I f_F = \frac{1}{4 [\la(s,t,u)]^2} \cdot \nn \\ 
&& \bigg[ - \lambda(s,t,u) ~ u_9
+3\,\left( 2\,s - t + 2\,u \right)~  u_{10}
   -  3\,\left( s - u \right)~  u_{11} \bigg] \\
&& \hspace{-7mm} F_3^A(s,u,t) f_I f'_F = \frac{1}{8 [\la(s,t,u)]^2} \cdot \nn \\ 
&& \bigg[(s-t+3u) \lambda(s,t,u)~ u_3 + 3 (2s-t+2u)(s-t+3u) ~ u_4 \nn \\
&& -3\left(s- t+ 3\,u \right) \left(s- u \right)~  u_5  
   - \left( 5\,{s^2} - 7\,s\,t + 2\,{t^2} + 2\,s\,u - t\,u - 
     7\,{u^2} \right)~ u_6  \nn \\ 
&& +\left(2s^2-st-t^2-4su+5tu+2u^2 \right)~ u_7-i\lambda(s,t,u) ~ u_{12}\bigg] \\
&& \hspace{-7mm} F_3^A(s,u,t) f'_I f_F = \frac{1}{8 [\la(s,t,u)]^2} \cdot \nn \\
&& \bigg[(3s-t+u) \lambda(s,t,u) ~ u_3 +3(3s-t+u) (2s-t+2u)~ u_4 \nn \\
&& - 3\,\left( 3\,s - t + u \right) \,\left( s - u \right)~ u_5 
 - \left( 7\,{s^2} + s\,t - 2\,{t^2} - 2\,s\,u + 7\,t\,u - 
     5\,{u^2} \right)~  u_6 \nn \\ 
&& +\left(2s^2+5st-t^2-4su-tu+2\,{u^2} \right)~  u_7
             - i \lambda (s,t,u) ~ u_{12} \bigg]\\ 
&& \hspace{-7mm} F_3^A(s,u,t) f'_I f'_F = \frac{1}{8 [\la(s,t,u)]^2} \cdot \nn \\ 
&& \bigg[ -2(2s-t+2u) \lambda(s,t,u)~ u_1+2(s-u) \lambda(s,t,u)~ u_2
    + (s-u) \lambda(s,t,u)~ u_8 \nn \\
&& -(s+u) \lambda(s,t,u)~ u_9 
  +\left(5\,{s^2}- 7\,s\,t+ 2\,{t^2}+ 14\,s\,u - 7\,t\,u 
 + 5\,{u^2} \right)~  u_{10} \nn \\ 
&&  -3 \left(s- t+ u \right) \,\left( s - u \right) ~  u_{11} \bigg] ~ .  
\eeqa

\bigskip

{\bf Appendix 2 . Form Factors in the $f'_I f_F$  Structure  }

For the sum rules based on $f'_If_F$ structures (see Eq.(\ref{unk}))
we obtain the general results 
\beq 
\label{general2}
F^A_1(t) =-\, F^V_1(t); \; F^A_2(t)= F^V_2(t)\; ,
\eeq
and only the perturbative diagram contributes to $F^{V(A)}_2$. Since the 
form factors
$F_3^{V(A)}$ contribute to the semileptonic decays at ${\cal{O}}(m_\ell^2/
q^2)$ \cite{Kor94} and are thus difficult to measure, we will not present 
results for these quantities here.

The stability behavior of the sum rules as a function of the Borel mass 
is here  very similar to that shown in Figs. 7,9 and 12, and the pole 
fits also describe very well the $t$-dependence of the form factors 
(as well as the  
curves shown in Figs.~8,10 and 13). In Table IV
we give the pole parametrization of the form factors for the process 
$\La_b\to\La_c \ell\nu_\ell$ for two values
of $\ka$  (see Eq. (\ref{cond})) and for the two forms of regions of 
continuum contribution considered. The other parameters are exactly the 
same as discussed in section IV. The continuum thresholds are obtained
using method 1 for $\Lambda_b$ and $\Lambda_c$, and method 2 for $\Lambda$.
\vskip1.0cm
\begin{center}
{\bf Table IV}

{\small Pole parametrization of the form factors for the process 
$\La_b\to\La_c \ell\nu_\ell$ for different values of the 
parameters and choices of continuum model.}

\vskip0.5cm
\begin{tabular}{c|c|c|c}\hline
continuum model & $\;\ka\;$ &$\;F_1^V(t)\;$ & $F_2^V(t)(\GeV^{-1})$\\
\hline
rectangular & 1 &${7.73/(17.79-t)}$ &${-0.20/(13.51-t)}$ \\
rectangular &2 &${8.93/(18.70-t)}$ &${-0.22/(14.91-t)}$\\
triangular&1  & ${16.60/(24.24-t)}$ &${-0.55/(18.63-t)}$\\
triangular&2 & ${19.02/(25.64-t)}$ &${-0.61/(19.10-t)}$\\
\hline
\end{tabular}
\end{center}
\vskip1.0cm

  We can see from Eq.~(\ref{FG}) that the IW function at
${\cal{O}}(1/m)$ is given by $G_1^A=-F_1^A-(M_{\La_F}-M_{\La_I})F_2^A$.
The extrapolation of the fits given in Table IV to the maximal momentum 
transfer value yields $G_1^A(t_{\rm max})=1.0 ~  (0.93)$ for 
$\ka=2 ~  (1)$ for the
rectangular  region, and $G_1^A(t_{\rm max})=1.04~  (1.03)$ for 
$\ka=2~  (1)$ for the
triangular region, in very good agreement with the HQET prediction.

With the pole parametrizations given above,
the $\Lambda_b\rightarrow\Lambda_c\ell\nu_\ell$  decay rate is
\beq
\Gamma(\La_b\to \La_c\ell^-\bar\nu_{\ell})=(5.1 \pm 1.8) \times 10^{-14} 
\; {\rm GeV} ~ ,
\label{deb2}
\eeq
where the errors reflect variations of $\kappa$ from 1 to 2 and the 
different choices for the continuum contribution. This value is 
larger than 
the one obtained based on $f_I$ and $f_F$ structures (see Eq.~(\ref{deb1})),
but it is still in agreement with other 
predictions \cite{Kor94,ILKK97,DHHL96,lls98} and with the 
experimental  upper limit \cite{PDG98} (see Eq.~(\ref{upb})).

In the case of the process $\Lambda_c\rightarrow\Lambda\ell\nu_\ell$,
 when the form factors are extracted from the sum rules based on 
$f'_I f_F$ structure, we obtain a  very large decay rate
\beq
\Gamma(\La_c^+\to \La + e^+ +  \nu_e) = (19 \pm 5) \times 10^{-14} 
\; {\rm GeV} ~, 
\label{dec2}
\eeq
which is much larger  than the experimental value\cite{PDG98}
(see Eq.~(\ref{upb})) and more than two times larger than the value
obtained using the sum rules based on $f_I f_F$ structure
(see Eq.~(\ref{dec1})). The mathematical reason for this discrepancy
is that while the $F_1^{V(A)}$ form factors extracted from both
$f_If_F$ and $f'_If_F$ structures are in very good agreement with 
each other, this is not 
the case for the $F_2^{V(A)}$ form factors, which are zero  when the
$f_If_F$ structure is used and of about  $-0.1(-0.2)$ (for $\ka=1(2)$) at
$t=0$ for the $f'_If_F$ structure. This can be seen in Table V, 
where
we give the pole parametrization of the form factors for the process 
$\La_c\to\La \ell\nu_\ell$ extracted from
the sum rules based on the $f'_I f_F$ structure.

\eject
\begin{center}
{\bf Table V}

{\small Pole parametrization of the form factors for the process 
$\La_c\to\La \ell\nu_\ell$ for different values of the 
parameters and choices of the continuum model.}

\vskip0.5cm
\begin{tabular}{c|c|c|c}\hline
continuum model & $\;\ka\;$ &$\;F_1^V(t)\;$ & $F_2^V(t)(\GeV^{-1})$\\
\hline
rectangular& 1 &${1.46/(2.98-t)}$ &${-0.19/(1.80-t)}$ \\
rectangular&2 &${1.78/(3.52-t)}$ &${-0.21/(1.94-t)}$\\
triangular&1  & ${2.80/(4.10-t)}$ &${-0.55/(2.84-t)}$\\
triangular&2 & ${3.06/(4.59-t)}$ &${-0.53/(3.00-t)}$\\
\hline
\end{tabular}
\end{center}
\vskip1.0cm

The contribution of $F_2^{V(A)}$ is responsible for the strong differences 
in the results for the calculated decay rates. In the 
$\Lambda_b\rightarrow\Lambda_c\ell\nu_\ell$ case, an inspection
of Tables IV and V shows that although the $F_1^{V(A)}$ form factors 
are of the same order of magnitude in the $\La_b$ and $\La_c$ decays, the
$F_2^{V(A)}$ form factors are one order of magnitude larger in the
$\La_c$ decay than in the $\La_b$ decay. The 
results for the $\La_b$ decay did not change  much because 
the values of the form factors $F_2^{V(A)}$ are consistent with
zero for both $f_If_F$ and $f'_If_F$ structures.

\bigskip

{\bf Appendix 3 . Form Factors in the $f_If'_F$  Structure }

The observed discrepancy in the results for $F_2^{V(A)}$ form factors 
comparing the $f_If_F$ and $f'_If_F$ structures for the 
$\La_c\to\La \ell\nu_\ell$ process described above leads us now to 
consider the form factors obtained through the
$f_If'_F$ structure. For these sum rules we obtain 
\beq 
\label{general3}
F^A_1(t) =-\, F^V_1(t); \; F^A_2(t)= -F^V_2(t)\; .
\eeq
Comparing Eq.~(\ref{general3}) with Eq.~(\ref{general2}) we see that the 
general results are correct only if $F_2^{V(A)}=0$, which is true
only for the  $f_If_F$ structure. Therefore, we can foresee more
discrepancies here. For the $F_1$ form factor, this structure gives a result
about 30\% smaller than in the previous studied structures for the
$\La_b$ decay, but gives a result with a factor more than two smaller for
the $\La_c$ decay. Moreover, while it is still possible to parametrize
the $t$-dependence of $F_1$ with a monopole form in the case of the $\La_b$ 
decay (with an agreement similar to the other structures), this is not 
possible for the $\La_c$ decay. The pole mass
needed to fit the $t$ behavior is smaller than  
$t_{\rm max}=(M_{\La_c}-M_{\La})^2$. We have used an exponential form 
factor of the type  $\exp{(t-a)/b}$
to fit the $t$-dependence of $F_1$, but the agreement is rather poor.

For the $F_2$ form factor a monopole fit is possible for both decays,
and its magnitude is about  20\% smaller than for the structure $f'_If_F$
for the $\La_b$ decay (therefore, consistent with zero), and a factor 
more than two smaller for the $\La_c$ decay.

In Table VI we give the pole parametrization of the form factors for the 
two processes $\La_b\to\La_c \ell\nu_\ell$ and 
$\La_c\to\La \ell\nu_\ell$, extracted from
the sum rules based on the $f_If'_F$ structure.
\vskip1.0cm
\begin{center}
{\bf Table VI}

{\small Parametrization of the form factors for the processes 
$\La_b\to\La_c \ell\nu_\ell$ and $\La_c\to\La \ell\nu_\ell$ for different 
values of the parameters and continuum models.}

\vskip0.5cm
\begin{tabular}{c|c|c|c|c|c}\hline
\multicolumn{2}{c|}{}&\multicolumn{2}{|c}
{$\La_b\to\La_c \ell\nu_\ell$} & \multicolumn{2}{|c}
{$\La_c\to\La \ell\nu_\ell$}\\
\hline
continuum model & $\;\ka\;$ &$\;F_1^V(t)\;$ & $F_2^V(t)(\GeV^{-1})$
&$\;F_1^V(t)\;$ & $F_2^V(t)(\GeV^{-1})$\\
\hline
rectangular& 1 &${6.66/(20.27-t)}$ &${-0.21/(15.15-t)}$ 
&$e^{(t-2.07)/0.95}$ &${-0.15/(2.85-t)}$ \\
rectangular&2 &${8.13/(22.50-t)}$ &${-0.22/(13.63-t)}$
&$e^{(t-2.08)/0.97}$ &${-0.17/(3.38-t)}$ \\
triangular&1  & ${13.74/(26.68-t)}$ &${-0.41/(18.65-t)}$
&$e^{(t-2.03)/1.09}$ &${-0.18/(2.79-t)}$ \\
triangular&2 & ${16.17/(29.12-t)}$ &${-0.45/(19.04-t)}$
&$e^{(t-1.72)/1.02}$ &${-0.20/(3.05-t)}$ \\
\hline
\end{tabular}
\end{center}
\vskip1.0cm

With the above parametrizations we obtain for the
the decay rates 
\beq
\Gamma(\La_b\to \La_c\ell^-\bar\nu_{\ell}) = (1.8 \pm 0.8) \times 10^{-14} 
\; {\rm GeV},
\label{deb3}
\eeq
and 
\beq
\Gamma(\La_c^+\to \La + e^+ +  \nu_e) = (2 \pm 1) \times 10^{-14} 
\; {\rm GeV}.
\label{dec3}
\eeq
Therefore, the sum rules based on the structure $f_If'_F$ give too small
decay rates for both $\La_b$ and $\La_c$  decays. Of course the variation 
in the three structures considered is much larger in the case of the 
$\La_c$ decay 
and this should be expected since the simple current used is not the
best suited to describe $\La$, as can be seen by the short-dashed line in
Fig. 11, which shows  that the {\bf 1} structure
is not trustworth in the mass sum rule, and this structure is related
with the $f'$ structure in the decay.

It is important to mention that even with this structure
the extrapolation of the fits given in Table VI to the maximal momentum 
transfer in the $\La_b$ decay yields for the IW  function the 
value $G_1^A(t_{\rm max})=0.97\pm0.04$  (where 
$G_1^A=-F_1^A-(M_{\La_F}-M_{\La_I})F_2^A$),
 in very good agreement with the HQET prediction.

\end{document}